\documentclass[11pt]{article}
\usepackage[margin=1in]{geometry}

\usepackage[T1]{fontenc}
\usepackage{lmodern}
\usepackage{microtype}
\usepackage{amsmath,amssymb,amsfonts,latexsym,amsthm}
\usepackage{color,graphicx}
\usepackage{url} 
\usepackage{fancybox}
\usepackage{algorithm}
\usepackage[noend]{algpseudocode}
\usepackage{mathtools,thmtools}
\usepackage{enumerate,xspace}
\usepackage{siunitx} 
\usepackage[commandnameprefix=ifneeded]{changes}
\usepackage{xcolor}
\usepackage{comment}
\usepackage[hidelinks]{hyperref}

\declaretheorem[name=Theorem]{theorem}
\declaretheorem[name=Lemma, sibling=theorem]{lemma}
\declaretheorem[name=Corollary, sibling=theorem]{corollary}
\declaretheorem[name=Definition, sibling=theorem]{definition}
\declaretheorem[name=Observation, sibling=theorem]{observation}
\declaretheorem[name=Fact, sibling=theorem]{fact}

\newcommand{\eps}{\varepsilon}
\renewcommand{\epsilon}{\eps}

\renewcommand{\leq}{\leqslant}
\renewcommand{\geq}{\geqslant}
\newcommand{\dG}{d_G}
\newcommand{\dH}{d_H}
\newcommand{\dE}{d_E}

\theoremstyle{plain}
\newenvironment{myquote}%
  {\list{}{\leftmargin=4mm\rightmargin=4mm}\item[]}%
  {\endlist}
\newenvironment{claiminproof}{\begin{myquote}\noindent\emph{Claim.}}{\end{myquote}}
\newenvironment{proofinproof}{\begin{myquote}\noindent\emph{Proof.}}{\hfill $\lhd$ \end{myquote}}

\newcommand{\B}{\ensuremath{\mathcal{B}}}

\newcommand{\REAL}{\ensuremath{\mathbb{R}}}
\newcommand{\Reals}{\REAL}

\DeclareMathOperator{\diam}{diam}

\DeclareMathOperator{\dist}{dist}

\DeclarePairedDelimiter\ceil{\lceil}{\rceil}

\newcommand{\myomit}[1]{}

\newcommand{\stub}{\mathrm{stub}}

\title{A Well-Separated Pair Decomposition of Linear Size\\ for Unit-Disk Graphs}
\author{
  Mark de Berg\thanks{Department of Mathematics and Computer Science, TU Eindhoven, the Netherlands, \texttt{M.T.d.Berg@tue.nl}. Supported by the Dutch Research Council (NWO) through Gravitation-grant NETWORKS-024.002.003. Part of the work was done while he was visiting IISc Bengaluru as Rukmini-Gopalakrishnachar Chair.}
  \and
  Joachim Gudmundsson\thanks{School of Computer Science, University of Sydney, Australia, \texttt{joachim.gudmundsson@sydney.edu.au}.}
  \and
  Sampson Wong\thanks{Department of Computer Science, University of Copenhagen, Denmark, \texttt{sawo@di.ku.dk}. Supported by the European Union's Marie Sk{\l}odowska-Curie Actions Postdoctoral Fellowship, grant 101146276.}
}
\date{}

\begin{document}

\maketitle

\begin{abstract}
We prove that every connected Euclidean-weighted unit-disk graph on $n$ vertices has 
a $(1/\eps)$-well-separated pair decomposition of size $O(n\eps^{-2}\log(1/\eps))$. 
This resolves an open problem posed by Gao and Zhang in~2003. 
We also show that this bound is tight: there are connected unit-disk graphs for which any
$(1/\eps)$-well-separated pair decomposition has size~$\Omega(n\eps^{-2}\log (1/\eps))$. 
\end{abstract}


\section{Introduction}\label{sec:introduction}

\subparagraph{Background.}
Let $(V,\dist)$ be metric space on a set $V$of $n$ points. 
We say that a pair $(A,B)$ of subsets of $V$
is \emph{well separated} if the diameters of the sets $A$ and $B$ are relatively small compared to
the distance between the sets. More precisely, $A$ and $B$ are
\emph{$(1/\eps)$-well-separated}, for some $0\leq \eps\leq 1$,
if $\max\left( \diam(A),\diam(B) \right) \leq \eps\cdot \dist(A,B)$, 
where $\diam(\cdot)$ denotes the diameter of a set and $\dist(A,B) := \min \{ \dist(p,q) : p\in A \mbox{ and } q\in B\}$. 
Thus, the distances between the points in $A$ 
and the points in~$B$ are the same up to a multiplicative factor of $(1+\eps)$.
A \emph{$(1/\eps)$-well-separated pair decomposition}, or \emph{$(1/\eps)$-WSPD} for short, of~$V$ 
is a collection~$\Psi := \{ (A_i,B_i) : 1\leq i\leq |\Psi| \}$
of pairs of subsets of~$V$ such that (i)~every pair $(A_i,B_i)$ is $(1/\eps)$-well-separated
and (ii)~for each $(p,q)\in V\times V$ there is a unique pair $(A_i,B_i)\in \Psi$
with $p\in A_i$ and $q\in B_i$ (or vice versa). The \emph{size} of a WSPD is the number
of pairs it consists of.

Callahan and Kosaraju~\cite{callahan-kosaraju} introduced the WSPD for point sets
in Euclidean space and they showed that any set $V$ of $n$ points in $\Reals^d$ admits
a $(1/\eps)$-WSPD of size~$O(n/\eps^d)$, for any fixed~$\eps>0$. The WSPD has become a fundamental tool in computational geometry, with applications to
spanners, nearest neighbours, distance oracles, diameter approximation, and many other
proximity problems~\cite{DBLP:conf/stoc/AryaDMSS95,DBLP:conf/soda/AryaM02,
DBLP:conf/soda/CallahanK93,DBLP:journals/siamcomp/GaoZ05,
DBLP:conf/compgeom/Har-Peled01,DBLP:journals/siamcomp/Har-PeledM06,
DBLP:journals/siamcomp/NarasimhanS00,DBLP:books/daglib/0017763,
DBLP:books/tf/18/Smid18,DBLP:conf/sigmod/WangY0S21}.

The versatility of the WSPD in Euclidean spaces naturally raises the question of which other 
metric spaces admit small WSPDs.
Clearly, not every metric space admits a $(1/\eps)$-WSPD of subquadratic size for any fixed~$\eps>0$. 
For example, consider a star graph\footnote{Here and in the sequel,
we assume that the given graph is equipped with the shortest-path metric.} with $n$ vertices, where each edge of the 
star has weight~1. Then any $(1/\eps)$-WSPD must have quadratic size when $\eps<1$; in fact,
the collection $\{ (\{u\},\{v\}) : u,v \in V\}$ is the only $(1/\eps)$-WSPD when $\eps<1$.
Fortunately, several other metrics \emph{do} admit small $(1/\eps)$-well-separated WSPDs. For example, 
WSPDs of size
$O_\eps(n)$  have been constructed for doubling 
metrics~\cite{DBLP:journals/corr/abs-2601-22728,DBLP:journals/siamcomp/Har-PeledM06,DBLP:conf/stoc/Talwar04} 
and for $c$-packed graphs~\cite{DBLP:conf/wads/DeryckereGRSW25},
whereas WSPDs of size $O_\eps(n \log n)$ have been constructed for so-called low-density graphs~\cite{DBLP:conf/soda/GudmundssonW26}. 
Here, the $O_\eps(\cdot)$ notation hides factors that depend only on~$\eps$; 
in other words, the asymptotic bound assumes that $\eps$ is a fixed constant. %

In this paper, we focus on WSPDs in unit-disk graphs. A \emph{unit-disk} graph is a graph $G=(V,E)$ where
$V$ is a set of points in the Euclidean plane and $E := \{ (u,v) : |uv|\leq 1\}$; here, $|uv|$ denotes
the length of the segment~$uv$.
In other words, $G$ contains an edge between two of its vertices $u$ and $v$ if and only if the disk of diameter~1 centred at $u$ intersects the disk of diameter~1 centred at~$v$.
We consider \emph{weighted unit-disk graphs}, where the weight of an edge $(u,v)\in E$
is equal to $|uv|$. Unit-disk graphs were introduced as a model for communication networks 
such as wireless ad hoc networks, and they have been studied extensively in distributed 
algorithms---see for example the edited volume by Wagner and Wattenhofer~\cite{WagnerWattenhofer2007} 
and the book by Li~\cite{Li2008}. They have also been studied widely in computational geometry.
For example, there are papers on single-source shortest paths~\cite{DBLP:journals/cgt/BrewerW26,DBLP:journals/comgeo/CabelloJ15,
DBLP:journals/dcg/KaplanMRSS20,DBLP:journals/siamcomp/Liu22,DBLP:journals/dcg/WangX20}, approximate distance oracles~\cite{DBLP:journals/jocg/ChanS19a,DBLP:conf/stoc/GaoZ03},
routing~\cite{DBLP:journals/algorithmica/KaplanMRS18,DBLP:conf/isaac/MulzerW20},
and reverse shortest paths~\cite{DBLP:conf/compgeom/ChanH25a,DBLP:conf/esa/KaplanKSS23,
DBLP:journals/jocg/0001Z23} in weighted unit-disk
graphs.
Many of these classic graph problems have also been studied in unweighted unit-disk graphs and (arbitrary) disk graphs. 

Given that WSPDs are a fundamental tool to design efficient algorithms and data structures,
and given the importance of unit-disk graphs, it is not surprising that WSPDs for unit-disk graphs have already been studied before.
In 2003, Gao and Zhang~\cite{DBLP:conf/stoc/GaoZ03} (see also the journal version~\cite{DBLP:journals/siamcomp/GaoZ05}) showed how to construct a $(1/\eps)$-WSPD of $O_{\eps}(n\log n)$ size.
In the conclusion of their paper, they state: \emph{The most notable open problem is the gap between $\Omega(n)$ and $O(n \log n)$ on the
number of pairs needed in the plane.}
In the decades that followed, this open problem has been restated a number of times~\cite{DBLP:reference/cg/2004,har-peled-raichel-robson,DBLP:conf/latin/KaplanMRS16,DBLP:phd/dnb/Seiferth16}, but despite this, little progress has been made. 
In 2025, Har-Peled, Raichel and Robson~\cite{har-peled-raichel-robson} improved the dependency on $\eps$ in the upper bound---the
bound by Gao and Zhang was $O(n \eps^{-4} \log n)$  while Har-Peled and Robson obtained a bound of
$O(n \eps^{-2} \log n)$---but the gap between $O_{\eps}(n \log n)$ and $\Omega_{\eps}(n)$ remains open to date. 

\subparagraph{Our result and techniques.}
We solve the open problem of Gao and Zhang by showing that any unit-disk graph admits a 
$(1/\eps)$-WSPD of size~$O_{\eps}(n)$. In fact, we completely settle the question of the 
required size of $(1/\eps)$-WSPDs for unit-disk graphs, including the dependency on~$\eps$ in
the bound: 
\begin{theorem}\label{thm:main_introduction}  
\begin{enumerate}[(i)]
\item Let $G$ be a connected Euclidean-weighted unit-disk graph on $n$ vertices. 
      For every sufficiently small $\eps>0$, the graph $G$ has a $(1/\eps)$-WSPD of size $O(n\eps^{-2} \log(1/\eps))$. Constructing the $(1/\eps)$-WSPD takes $O(n\log n+n\eps^{-2}\log(1/\eps))$ time.
\item For every $\eps>0$ and all sufficiently large $n$, there is a connected Euclidean-weighted unit-disk graph 
      on $n$ vertices for which every $(1/\eps)$-WSPD has size $\Omega(n\eps^{-2}\log(1/\eps))$.
\end{enumerate}
\end{theorem}

To prove the size bound in
Theorem~\ref{thm:main_introduction}(i), we build on the approach of
Har-Peled, Raichel, and Robson~\cite{har-peled-raichel-robson}. Their approach 
uses a hierachy of $r$-nets and then counts the number of nearby pairs of net vertices, 
at doubling distance scales. At any one scale,
their approach gives $O(n\eps^{-2})$ pairs,
and this bound can be tight. Applying their argument
separately at each of the $O(\log n)$ scales therefore gives only an
$O(n\eps^{-2}\log n)$ bound.

Our main technical contribution is a new charging argument that keeps
a set of vertex-disjoint paths, called \emph{stubs}, across all scales.
In the simple cases, packing arguments give a total of $O(n\eps^{-2})$
nearby pairs of net vertices. In the remaining case, for each nearby pair of net vertices, we choose a stub and we distribute one unit of charge to its edges. We show in our analysis that a stub edge can receive at most $O(\eps^{-2})$
total charge per level, and is only charged on $O(\log(1/\eps))$ levels. Since the stubs set contains at
most $n$ edges in total, this implies an upper bound of
$O(n\eps^{-2}\log(1/\eps))$ on the total charge. This improves the upper bound on the WSPD size from $O(n\eps^{-2}\log n)$ to $O(n\eps^{-2}\log(1/\eps))$; see Section~\ref{sec:upper-bound} for details.
\medskip

Next, we consider the lower bound in \autoref{thm:main_introduction}(ii). It is relatively simple to prove a lower bound of $\Omega(n\eps^{-2})$; if all vertices of~$G$ lie inside a disk of
diameter one, then the shortest path metric of~$G$ is the Euclidean metric, so a
scaled $\sqrt{n}\times\sqrt{n}$ grid already gives $\Omega(n\eps^{-2})$. To obtain the extra factor of
$\log(1/\eps)$, we must use the property of unit-disk graphs that two vertices can be close in the plane but far apart in the graph.

In our lower bound construction in Section~\ref{sec:LB}, the basic building block is an  \emph{antenna}. An antenna consists of one backbone and~$\Theta(1/\eps)$
teeth, where each tooth is a path of length~$\Theta(1/\eps)$. Vertices on neighbouring teeth can be close in the plane but far apart in the graph. We arrange the antennas in a hierarchy, so that far away from a vertex,
there are vertices from many antennas. The hierarchy has $\Theta(\log(1/\eps))$ levels,
and we show that each level requires $\Omega(n\eps^{-2})$ well-separated pairs. Since each well-separated pair contributes to only $O(1)$ levels, this gives
a total of $\Omega(n\eps^{-2}\log(1/\eps))$ well-separated pairs.
\medskip

Finally, in Section~\ref{sec:construction}, we prove the construction time in Theorem~\ref{thm:main_introduction}(i). Our construction time is optimal since we output $O(n\eps^{-2}\log(1/\eps))$ pairs, and there is an $\Omega(n \log n)$ lower bound by a reduction to the nearest pair problem. Our algorithm considers two kinds of pairs. Short pairs consist of sets of
Euclidean diameter at most one, and we obtain them by filtering a
Euclidean WSPD. Long pairs are produced by a recursion in the style of
Callahan and Kosaraju~\cite{callahan-kosaraju} on a hierarchy of weak nets. Short and long pairs require $O(n \log n + n\varepsilon^{-2})$ and $O(n \log n + n\eps^{-2}\log(1/\eps))$ time to construct, respectively.

\subparagraph{Partition vs. cover WSPD.}
Callahan and Kosaraju~\cite{callahan-kosaraju} define a WSPD to be a collection of well-separated pairs $(A_i,B_i)$
such that for every pair of points~$p,q\in V$ there is a unique pair $(A_i,B_i)$ such that
$(p,q)\in A_i\times B_i$ (or vice versa). Recently, Har-Peled et al.~\cite{har2011geometric,DBLP:journals/siamcomp/Har-PeledM06,har-peled-raichel-robson} proposed a relaxed definition
where the pair $(A_i,B_i)$ is not necessarily unique. We call the unique variant a partition WSPD, and the non-unique variant a cover WSPD.

Theorem~\ref{thm:main_introduction} holds for both partition and cover WSPDs. Sections~\ref{sec:upper-bound},~\ref{sec:LB} and~\ref{sec:construction} together prove Theorem~\ref{thm:main_introduction} for the cover variant, although the lower bound in Section~\ref{sec:LB} applies to both the cover and partition variants. In Section~\ref{sec:partition}, we refine the cover WSPD into a partition WSPD. The WSPD for both variants have the same asymptotical size and running time.

\subparagraph{Application to approximate distance oracles.}
One of the standard applications of WSPDs is to obtain an efficient \emph{$(1+\varepsilon)$-approximate distance oracle},
which stores a given graph~$G$ in such a way that one can quickly report, for given query vertices $u$ and~$v$,
a value $d^*$ such that $\dist_G(u,v)\leq d^* \leq (1+\eps)\cdot \dist_G(u,v)$.
Approximate distance oracles have been studied for
planar graphs~\cite{chan-skrepetos2019planar,klein2002planar,
le2023planar,DBLP:conf/focs/LeW21,thorup2004planar,
wulffnilsen2016planar},
minor-free graphs~\cite{DBLP:conf/soda/ChangCLMST24,
kawarabayashi-klein-sommer2011},
geometric spanners~\cite{DBLP:journals/talg/GudmundssonLNS08},
$c$-packed graphs~\cite{deryckere2025cpacked},
low-density graphs~\cite{DBLP:conf/soda/GudmundssonW26},
and doubling metrics~\cite{DBLP:journals/siamcomp/Har-PeledM06,
talwar2004doubling}.

Gao and Zhang~~\cite{DBLP:conf/stoc/GaoZ03} and Chan and Skrepetos~\cite{chan-skrepetos} 
provided an approximate distance oracle for unit-disk graphs that uses $O_\eps(n\log n)$ storage and has $O(1)$ query time.
As shown in Section~\ref{sec:distance_oracle}, our new WSPD improves the storage usage of this oracle to $O_\eps(n)$.
\begin{restatable}{theorem}{DistanceOracle}  \label{thm:oracle}
Given a connected Euclidean-weighted unit-disk graph on $n$ vertices, 
and $0 < \varepsilon \leq 1/8$, there exists a $(1+\varepsilon)$-approximate distance oracle 
with $O(1)$ query time and that uses $O(n \varepsilon^{-2}\log(1/\varepsilon))$ storage.
The oracle can be constructed in $O(n \varepsilon^{-5} \log^3 n + n \varepsilon^{-6} \log(1/\varepsilon))$ expected  time. 
\end{restatable}

\section{The upper bound}\label{sec:upper-bound}

Let $V$ be a set of $n$ points in the plane. Let $G=(V,E)$ be the corresponding unit-disk
graph. Recall that the edges of~$G$ are given by $E := \{ (u,v) : |uv|\leq 1\}$. We denote the shortest-path distance in $G$ between any two vertices~$u,v\in V$
by~$\dist_G(u,v)$, and the diameter of a subset $A\subset V$ by $\diam_G(A) := \max_{u,v\in A} \dist_G(u,v)$.
Instead of $\diam_G(V)$, we will sometimes write $\diam(G)$. 

Below we first discuss so-called \emph{$r$-nets}, and we state a result by
Har-Peled, Raichel, and Robson~\cite{har-peled-raichel-robson} who show that a small
WSPD for a unit-disk graph can be constructed from a hierarchy of $r$-nets.
After that we explain our new construction.

\subsection{The Har-Peled--Raichel--Robson reduction} \label{sec:reduction}
Recall that an $r$-net of a graph $G$ is defined as follows~\cite{DBLP:journals/siamcomp/Har-PeledM06}.
\begin{definition}[$r$-net]
\label{definition:net}
An $r$-net of a graph~$G=(V,E)$ is a set $N\subseteq V$ of vertices such that
\begin{enumerate}[(i)]
\item $N$ is \emph{$r$-separated}: $\dist_G(x,y) \geq r$ for any two distinct vertices $x,y\in N$, 
\item $N$ is \emph{$r$-covering}: for every vertex $u\in V$ there is a vertex $x\in N$ such that $\dist_G(u,x)\leq r$.
\end{enumerate}
\end{definition}

\noindent The WSPD construction of Har-Peled, Raichel, and Robson~\cite{har-peled-raichel-robson}  
works with hierarchies of $r$-nets, as defined next.
\begin{definition}[Hierarchy of $r$-nets]\label{def:net-hierarchy}
A hierarchy of $r$-nets of a graph~$G$ is a nested sequence $N_0\supseteq N_1\supseteq\cdots\supseteq N_L$
in which each $N_i$ is an $r_i$-net of~$G$, where $r_0>0$ is a fixed constant, $r_i=2^i r_0$,   
and $L := \ceil{\log (\diam(G)/r_0)}$. We refer to $i$ as the \emph{level} of~$N_i$.
\end{definition}
One can construct a hierarchy of $r$-nets for a given graph~$G$ as follows.
First, take an arbitrary vertex $x\in V$ and set  $N_L:= \{x\}$. Note that the definition
of~$L$ implies that $r_L \geq \diam(G)$, so the set~$\{x\}$ is a valid $r_L$-net.
Then, for $i = L-1, \ldots, 0$, inductively construct $N_{i+1}$ from $N_i$ in a greedy manner.
To this end, initialise $N_i := N_{i+1}$ and then, as long as there is a vertex $v\not\in N_i$ with
$\dist(v,N_i) >r_i$, add that vertex to~$N_i$. When no more vertices can be added,
$N_i$ is an inclusion-maximal $r_i$-separated set.
The top-down construction ensures the sequence $N_i$ is nested, and maximality 
under the $r_i$-separated property implies that $N_i$ is $r_i$-covering.
\medskip

Har-Peled, Raichel and Robson show how to obtain a WSPD from a hierarchy of $r$-nets,
and they prove the following about size of the resulting WSPD.
\begin{fact}[Har-Peled, Raichel, and Robson~\cite{har-peled-raichel-robson}]\label{fact:hpr}
Let $G$ be a connected Euclidean-weighted unit-disk graph on $n$ vertices, and let 
$N_0,\ldots,N_L$ be a hierarchy of $r$-nets with $r_0=12$.  For each $i$, let $m_i$ be 
the number of pairs $x,y\in N_i$ such that $\dist_G(x,y) \leq 32r_i/\eps$. 
Then $G$ has a $(1/\eps)$-WSPD of size
$O\Big(n\eps^{-2}+\sum_{i=0}^{L}m_i\Big).$
\end{fact}
Har-Peled, Raichel, and Robson~\cite{har-peled-raichel-robson}
then obtain a bound on $m_i$ through the following result.
\begin{fact}[Lemma 3.6 in~\cite{har-peled-raichel-robson}] \label{fact:packing}
    Let $G$ be a connected Euclidean-weighted unit-disk graph on $n$ vertices. 
    Let $r>8$ and let $N$ be a set %
    such that $\dist_G(x,y) \geq r$ for all $x,y\in N$.
    Then $|N| = O(n/r)$. Moreover, for any $x\in N$ and any $R\geq r$, the number of vertices 
    $y\in N$ such that $\dist_G(x,y) \leq R$ is $O(R^2/r)$.
\end{fact}
By applying Fact~\ref{fact:packing} with $r=r_i$ and $R = 32r_i/\eps$ to a given vertex $x \in N_i$, 
we can conclude that there are at most $O(r_i \eps^{-2})$ vertices in $N_i$ whose distance 
to vertex~$x$ is  $\leq 32r_i/\eps$. Summing this over all choices of $x$ thus gives $m_i = O(n \eps^{-2})$. 
\begin{observation}[Lemma 3.7 in~\cite{har-peled-raichel-robson}] \label{observation:neps2}
    For all $0 \leq i \leq L$, we have $m_i = O(n \eps^{-2})$.
\end{observation}
However, combining \autoref{fact:hpr} and \autoref{observation:neps2} only yields a bound of
$\sum_{i=0}^{L}m_i = O(n \eps^{-2} \log n)$ on the size of the WSPD. Indeed, we can have $L=\Theta(\log n)$,
since $R_L = \Theta(\diam(G))$ and $\diam(G)$ can be as large as~$n-1$. 
So there can be levels~$i$ where $m_i = \Theta(n/\eps^2)$. 
Therefore, in order to improve on the result by Har-Peled, Raichel, and Robson,
we need a global analysis that considers multiple levels simultaneously.

\subsection{Proof of the size bound in \autoref{thm:main_introduction}(i)}
\label{sec:large_scale}
Let $N_0,\ldots,N_L$ be a hierarchy of $r_i$-nets with $r_0=12$ and $r_i = 2^i r_0$.
We will prove that $\sum_{i=0}^{L}m_i = O(n \eps^{-2} \log (1/\eps))$. 
Together with \autoref{fact:hpr}, this proves our main result. 
It will be convenient to assume that $r_i \geq 768$, which is true for $i\geq 6$. 
Thus, from now on we focus on proving that $\sum_{i=6}^{L}m_i= O(n \eps^{-2} \log (1/\eps))$. 
This is sufficient since $\sum_{i=0}^{5}m_i = O(n \eps^{-2})$ by \autoref{observation:neps2}. 
We first introduce the main concepts needed to obtain our improved bound:
stubs, the stub set, and relevant pairs.

\subparagraph{Stubs, the stub set, and relevant pairs.} 
Our argument will be organised around short portions of shortest paths, which we will call stubs. 
They are defined as follows.
\begin{definition}[Stub]\label{def:stub}
Let $x$ be a vertex of the net~$N_i$, and assume that $|N_i|\geq 2$.  
A \emph{stub at $x$} is the prefix of a shortest path~$\pi(x,y)$ from $x$ to 
some other vertex~$y\in N_i$ that ends at the first vertex $u\in \pi(x,y)$
such that $\dist_G(x,u)\geq r_i/100$. 
\end{definition}
Note that stubs are well defined because~$\dist_G(x,y) \geq r_i$ (since $x,y\in N_i$).
Also note that a vertex $x\in N_i$ may define several different stubs, because there are $|N_i|-1$
choices for the vertex~$y$ and because the shortest path~$\pi(x,y)$ may not be unique.
For our purposes, the actual choice does not matter. Thus, to obtain a unique stub at~$x$
we pick an arbitrary vertex $y\in N_i$ and an arbitrary shortest path $\pi(x,y)$ t.
We denote the resulting stub by~$\stub(x)$ and we call $x$ the \emph{starting point} of~$\stub(x)$.

An important property of the stub of a vertex $x\in N_i$ is that its length, 
which equals $\dist_G(x,u)$, is between $r_i/100$ and $r_i/100 + 1$. 
This is true since the length of any edge---and,
in particular, the length of the last edge in the prefix---is at most~1. 
\medskip

\begin{figure}[tbh]
    \centering
    \includegraphics[scale=0.8]{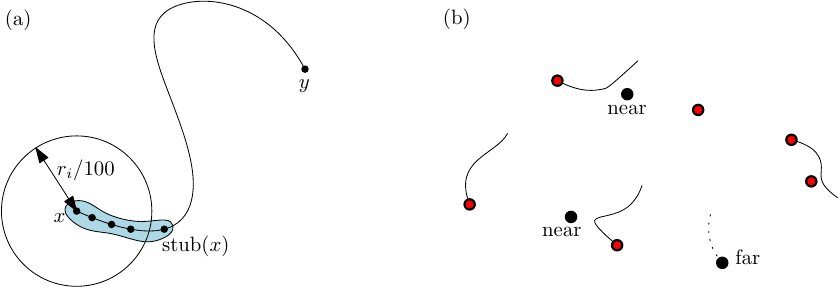}
    \caption{(a) Illustrating the stub of $x$. (b) The red vertices are in $N_{i+1}$ and the black vertices are in $N_i$. One of the black vertices is far from $S_{i+1}$ and two are near~$S_{i+1}$.}
    \label{fig:stub}
\end{figure}

\begin{figure}[tbh]
    \centering
    \includegraphics[scale=0.8]{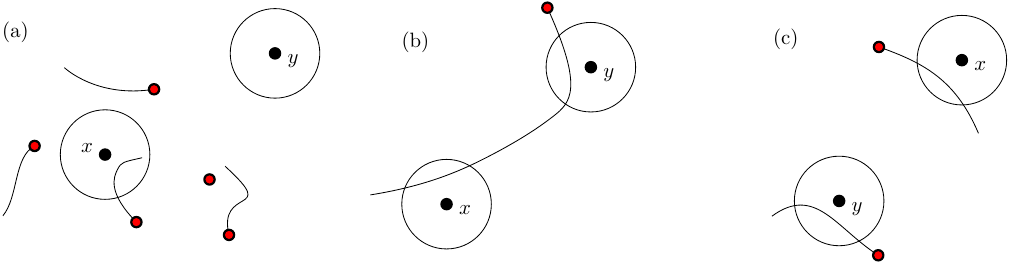}
    \caption{(a) Type~1 pairs, (b) Type~2 pairs, and (c) Type~3 pairs,.}
    \label{fig:stub types}
\end{figure}

Our goal will be to relate $\sum_{i=8}^L m_i$, the quantity we need to bound in order 
to apply \autoref{fact:hpr}, to vertices of stubs. To this end we will construct stubs 
across all levels to obtain a so-called stub set. Our charging argument relies on
the stubs in our stub set being pairwise disjoint. To ensure disjointness, we construct 
the stub set level by level, where we choose the starting points of the stubs at a certain
level to be relatively far from the previous stubs. Here the distance of vertex $x$ from a set~$S$
of stubs is the minimum distance from $x$ to any vertex in a stub in $S$; see Fig~\ref{fig:stub}.
More precisely, we define the subset as follows.
\begin{definition}[Stub set]\label{def:stub_set}
The \emph{stub set} $S_0$ is defined inductively.  Initially, $S_{L+1}=\emptyset$.  Then, for $i=L,L-1,\ldots,0$, suppose that $S_{i+1}$ has been defined.  Consider a net vertex $x\in N_i$. 
\begin{itemize}
  \item The vertex $x$ is \emph{far from $S_{i+1}$} if its distance from $S_{i+1}$ is
  greater than $8r_i/100$.
  \item The vertex $x$ is \emph{near $S_{i+1}$} if its distance from $S_{i+1}$ is
  at most $8r_i/100$. 
\end{itemize} 
Define $S_i$ to be $S_{i+1}$ together with the stubs for all net vertices of $N_i$
that are far from~$S_{i+1}$.
\end{definition}
The next lemma shows that the stubs in the stub set~$S_0$ are indeed pairwise disjoint.
\begin{lemma}[Disjointness]\label{lem:stub-disjoint}
The stub set $S_0$ is a set of pairwise disjoint stubs.
\end{lemma}
\begin{proof}
We first prove the lemma for stubs created at the same level, and then for stubs at different
levels.

Consider two stubs $\stub(x)$ and $\stub(y)$ created at the same level~$i$. 
Thus, $x,y\in N_i$ and so $\dist_G(x,y)\geq r_i$.
Recall that the length of the stubs at level~$i$ is at most $r_i/100 + 1$. Hence,
if $\stub(x)\cap\stub(y)\neq\emptyset$ then, by the triangle inequality, we would have 
$\dist_G(x,y) \leq 2(r_i/100 + 1) \leq r_i /40$; here the last inequality uses that $r_i > 768$.
But this would contradict $\dist_G(x,y)\geq r_i$. We conclude that $\stub(x)$ and $\stub(y)$ are disjoint.

Now consider two stubs $\stub(x)$ and $\stub(y)$ created at levels $i$ and $j$, respectively,
where $i < j$. Thus, when $\stub(x)$ is being created, $\stub(y)$ already exists. 
This implies that $\dist_G(x,\stub(y)) > 8 r_i/100$. Since the length of $\stub(x)$ 
is at most~$r_i/100+1$, we can conclude that $\dist_G(\stub(x),\stub(y)) \geq (8 r_i/100) - (r_i/100+1)>0$.
Hence, $\stub(x)$ and $\stub(y)$ are also disjoint when they are created at different levels.
\end{proof}
A consequence of disjointness is that the the total size of the stubs in the stub set~$S_0$ 
is linear. (Here the size of a stub is defined to be its number of edges, which is
the same as its number of vertices minus~1, since stubs are paths.)
\begin{corollary}[Linear size]
    \label{cor:stub_set_linear}
    Let $|P|$ denote the size of a stub~$P$. Then $\sum_{P\in S_0} |P| = O(n)$. 
\end{corollary}
Recall that we want to bound $\sum_{i=1}^L m_i$, where
$m_i$ is the number of pairs $x,y\in N_i$ such that $\dist_G(x,y) \leq 32r_i/\eps$.
To this end, we use the stub set~$S_0$ to classify such pairs of net vertices
into three types; see Fig~\ref{fig:stub types} for an illustration.
\begin{definition}[Relevant pairs and their types]\label{def:relevant}
A pair $\{x,y\}$ of distinct vertices of $N_i$ is \emph{relevant} if $\dist_G(x,y)\le {32r_i}/{\eps}$. 
Each relevant pair is classified as follows: 
\begin{itemize}
  \item $\{x,y\}$ has \emph{Type~1} if either $x$ or $y$ is far from~$S_{i+1}$.
  \item $\{x,y\}$ has \emph{Type~2} if both $x$ and $y$ are near~$S_{i+1}$ and their nearest stubs are the same.
  \item $\{x,y\}$ has \emph{Type~3} if both $x$ and $y$ are near~$S_{i+1}$ and their nearest stubs are different.
\end{itemize}
\end{definition}
The definition above refers to the nearest stubs of~$x$ and~$y$. Here,
ties between nearest stubs are broken arbitrarily (but once the nearest stub
of a vertex is chosen, it remains fixed).  
We define the level of a relevant pair $\{x,y\}$ to be the index~$i$ such that $x,y\in N_i$.
In the remainder of this section, we bound the total number of relevant pairs 
over all levels by separately bounding the numbers of pairs of Types~1,~2 and~3.

\subparagraph{Type~1 pairs.}
In the proof of the next lemma, we will show how to charge the Type~1 pairs
to the edges of the stub set, and argue that no edge is charged too many times.
\begin{lemma}[Type~1 pairs]\label{lem:Type_1}
There are $O(n\eps^{-2})$ Type~1 pairs in total over all levels~$i$.
\end{lemma}
\begin{proof}
Consider a Type~1 pair~$\{x,y\}$ of level~$i$ and
assume wlog that $x$ is far from~$S_{i+1}$. Definition~\ref{def:stub_set} thus implies that
$\stub(x)$ is a stub in our stub set. We charge the pair~$\{x,y\}$ 
evenly across all edges in $\stub(x)$, that is, each edge of $\stub(x)$
is charged a cost of $1/|\stub(x)|$, where $|\stub(x)|$ is the number
of edges of $\stub(x)$. This completes the description of the charging scheme.

Next, we bound the maximum amount of charge accumulated by an edge~$e$ in the stub set. 
Let $e$ be an edge of $\stub(x)$ for some $x\in N_i$. Then $e$ accumulates 
charge at level~$i$, and~$e$ accumulates no charge at any other level. 
For each Type~1 pair $\{x,y\}$ containing $x$, the charge accumulated by $e$ is at most $100/r_i$, 
since $|\stub(x)| \geq r_i/100$. The number of Type~1 pairs $\{x,y\}$ containing $x$ 
is upper bounded by the number of relevant pairs $\{x,y\}$ containing~$x$, 
which by Fact~\ref{fact:packing} is $O(r_i \eps^{-2})$. Putting this together, 
the edge~$e$ accumulates a charge of at most $O(\eps^{-2})$. 
Combining this with \autoref{cor:stub_set_linear}, the total charge accumulated 
by all edges in the stub set is $O(n\eps^{-2})$. 
Since each Type~1 pair distributes a cost of~1 in our charging scheme, 
we can conclude that there are $O(n \eps^{-2})$ Type~1 pairs.
\end{proof}

\subparagraph{Type~2 pairs.}
A Type~2 pair at level~$i$ consists of two vertices that are near~$S_{i+1}$
and whose nearest stubs are the same. 
In the next lemma, we will associate Type~2 pairs with their nearest stubs, 
and use a packing argument to prove that there cannot be too many Type~2 pairs 
associated to the same stub.
\begin{lemma}[Type~2 pairs]\label{lem:Type_2}
There are $O(n\eps^{-1})$ Type~2 pairs in total over all levels~$i$.
\end{lemma}
\begin{proof}
We will prove below that for any vertex $x\in N_i$ there are only
$O(1/\eps)$ vertices $y$ such that $\{x,y\}$ is a Type~2 pair. 
Since $|N_i|=O(n/r_i)$ by \autoref{fact:packing} and $r_i = 2^i r_0 = \Theta(2^i)$, 
it then follows that the total number of Type~2 pairs over all levels~$i$ is bounded by
\[
\sum_i |N_i| \cdot O(1/\eps) = O(n/\eps) \cdot \sum_i (1/r_i) = O(n/\eps),
\]
which is what we wanted to prove.

It remains to prove that any vertex $x\in N_i$ participates in $O(1/\eps)$ Type~2 pairs.
To this end, consider the set~$Y$ of all vertices~$y\in N_i$ such that $\{x,y\}$ is a Type~2 pair.
Let $P\in S_{i+1}$ be the nearest stub of~$x$ and observe that, by definition, $P$
is also the nearest stub for any $y\in Y$. Let $u_x$ be a vertex in $P$ that
is closest to~$x$. Similarly, for a vertex $y\in Y$, let $u_y$ be a vertex in $P$ that
is closest to~$y$. 
Since $x$ and the vertices $y\in Y$ are all near~$S_{i+1}$, \autoref{def:stub_set} implies that
\[
\dist_G(x,u_x) \leq 8r_i/100  \mbox{\ \ \ and \ \ \ } \dist_G(y,u_y) \leq 8r_i/100 \mbox{ for all $y\in Y$}.
\]
Thus, while $P$ itself can be quite long (since it comes from a level $j>i$),
the vertices $u_y$ are located on a subpath of~$P$ of length at most~$16 r_i/100$.
Now consider two distinct vertices $y,y'\in Y$. Since $Y\subseteq N_i$, we have $\dist_G(y,y')\geq r_i$ and so
\[
r_i \leq \dist_G(y,y') \leq \dist_G(y,u_y) + \dist_G(u_y,u_{y'}) + \dist_G(u_{y'},y')
                       \leq \dist_G(u_y,u_{y'}) + 16 r_i/100.
\]
Hence, 
\begin{equation}  \label{eq1}
\dist_G(u_y,u_{y'}) \geq 84 r_i/100.
\end{equation}
On the other hand, since $\{x,y\}$ is a relevant pair for any $y\in Y$,
we have $\dist_G(x,y) \leq 32 r_i/\eps$ by \autoref{def:relevant}. 
Thus, for all $y\in Y$ we have
\begin{equation} \label{eq2}
\dist_G(u_x,u_y) \leq \dist_G(u_x,x)+ \dist_G(x,y) + \dist_G(y,u_y) \leq (32r_i/\eps) + (16r_i/100).
\end{equation}
Recall that the stub~$P$ is a shortest path.  This implies that $\dist_G(u_x,u_y)=\dist_P(u_x,u_y)$
and $\dist_G(u_y,u_{y'})=\dist_P(u_y,u_{y'})$ for any two vertices $y,y'\in Y$. Combining this
with Inequalities~(\ref{eq1}) and~(\ref{eq2}) we conclude that 
$\dist_P(u_y,u_{y'}) = \Omega(r_i)$ for any two vertices $y,y'\in Y$
and $\dist_P(u_x,u_{y}) =O(r_i/\eps)$ for any $y\in Y$. Hence, $|Y|= O(1/\eps)$,
as claimed.
\end{proof}

\subparagraph{Type~3 pairs.}
A Type~3 pair at level~$i$ consists of two vertices that are near~$S_{i+1}$
and whose nearest stubs are different. 
The key insight is that these nearest stubs cannot both be much older than level~$i$. 
Then, we can charge the Type~3 pairs to the most recent of these two stubs. 
We formalise this argument in the following lemma.
\begin{lemma}[Type~3 pairs]\label{lem:Type_3}
There are $O(n\eps^{-2}\log(1/\eps))$ Type~3 pairs over all levels~$i$.
\end{lemma}
\begin{proof}
Consider a Type~3 pair $\{x,y\}$ at level~$i$. Let $P_x \neq P_y$ be their nearest stubs,
where $P_x$ was created at level~$j$ and $P_y$ was created at level~$k$.
Assume wlog that $i < j \leq k$. The following claim establishes that $P_x$ is
not much older than level~$i$.
\begin{claiminproof}
$j-i = O(\log(1/\eps))$. 
\end{claiminproof}
\begin{proofinproof}
Since $x$ and $y$ are a relevant pair, we have $\dist_G(x,y) \leq 32 r_i/\eps$ 
by \autoref{def:relevant}. Moreover, $\dist_G(x,P_x) \leq 8r_i/100$
and $\dist_G(y,P_y) \leq 8r_i/100$ since $x$ and $y$ are near~$S_{i+1}$. Hence, 
if we define~$c_1 := 33$ so that $32/\eps + 16/100 < c_1/\eps$ (since we
may assume $\eps<1$), then we have
\begin{equation} \label{eq:pxpy}
\dist_G(P_x,P_y) \leq \dist_G(P_x,x) + \dist_G(x,y) + \dist_G(y,P_y) < (c_1/\eps) r_i.
\end{equation}
Inequality~(\ref{eq:pxpy}) gives an upper bound on $\dist_G(P_x,P_y)$ in terms of~$r_i$.
To prove the claim, we now derive a lower bound on $\dist_G(P_x,P_y)$ in terms of~$r_j$.
In particular, we show 
\begin{equation} \label{eq:pxpy2}
\dist_G(P_x,P_y) > c_2 r_j  \mbox{ where $c_2 := 27/400$.} 
\end{equation}
We consider two cases.
\begin{itemize}
\item If $k=j$ (thus, $P_x$ and $P_y$ are both created at level~$j$) then starting points
      of $P_x$ and $P_y$ are at distance at least $r_j$ from each other. Since $P_x$ and $P_y$ have length
      at most~$r_j/80$, we therefore have $\dist_G(P_x,P_y) \geq r_j - 2 r_j/80 > c_2 r_j$.
\item If $k > j$, then $P_y$ already exists when $P_x$ is created. 
      By Definition~\ref{def:stub_set}, the starting point of $P_x$ is then at distance 
      greater than $8r_j/100$ from $P_y$. Since the length of $P_x$ is at most $r_j/80$, 
      we have $\dist_G(P_x,P_y) > 8r_j/100 - r_j/80 = c_2 r_j$.
\end{itemize}
Combining Inequalities~(\ref{eq:pxpy}) and~(\ref{eq:pxpy2}) we obtain
$(c_1/\eps) r_i > \dist_G(P_x,P_y) > c_2 r_j$, which implies that
$r_j/r_i < c_1/(c_2\eps)$. Since $r_j = 2^{j-i} r_i$, it follows that $j-i=O(\log(1/\eps))$.
\end{proofinproof}
We proceed with a charging scheme similar to \autoref{lem:Type_1}. 
In particular, we charge the Type~3 pair $\{x,y\}$ to stub~$P_x$,
by charging a cost of $1/|P_x|$ to each edge of~$P_x$.
We first bound the number of times a stub~$P$ can be charged. 
\begin{claiminproof}
Any stub $P$ created at level $j$ is charged $O(r_j \eps^{-2} \log (1/\eps))$ times.    
\end{claiminproof}
\begin{proofinproof}
Fix a level~$i<j$.
We first bound the number of vertices $x \in N_i$ such that some pair $\{x,y\}$ 
charges to~$P$. Every such pair $\{x,y\}$ has a vertex~$x$ that is near~$S_{i+1}$
with $P$ as its nearest stub. Let $u_x$ be a vertex of $P$ that is closest to~$x$. 
By Definition~\ref{def:stub_set}, we have $\dist(x,u_x) \leq 8r_i/100$. 
On the other hand, any two net vertices $x, x' \in N_i$ are at distance at least~$r_i$
from each other. Hence, by the triangle inequality, 
$\dist(u_x, u_{x'}) \geq r_i-2(8r_i/100)=84r_i/100$. 
Since the length of $P$ is at most $r_j/80$, we may conclude that there 
are $O(r_j/r_i)$ vertices $x\in N_i$ that may define a pair $\{x,y\}$
that charges to~$P$.
\smallskip

Now fix a vertex~$x\in N_i$. We will bound the number of Type~3 pairs~$\{x,y\}$ 
that charge to $P$. Because $\{x,y\}$ is relevant, we know that $\dist_G(x,y)\leq 32r_i/\eps$ from $x$. 
Since $N_i$ is $r_i$-separated, Fact~\ref{fact:packing} states that there 
are $O((32r_i/\eps)^2/r_i)=O(r_i\eps^{-2})$ vertices~$y$ within this distance from~$x$. 
In other words, there are $O(r_i\eps^{-2})$ pairs $\{x,y\}$ that charge to $P$, for a fixed~$x$. 
\smallskip

Putting this together, at a fixed level~$i$, a stub $P$ created at some level~$j>i$
is charged only $O((r_j/r_i)\times (r_i \eps^{-2}) = O(r_j \eps^{-2})$ times. 
By the Claim above, we know that $j-i= O(\log(1/\eps))$ when $P$ is charged from level~$i$.
Hence, $P$ is charged $O(r_j \eps^{-2} \log (1/\eps)$ times in total.   
\end{proofinproof}
Finally, we will bound the total charge, which is equal to the number of
Type~3 pairs, by summing the costs assigned to the edges of the stubs.
Let $e$ be such an edge and assume it belongs to a stub~$P$ created at level~$j$.
We argued above that $P$ is charged only $O(r_j \eps^{-2} \log (1/\eps)$ times.
Since $P$ has at least $r_j/100$ edges and the costs of a single charge are distributed
evenly, the edge $e$ is charged a cost of at most $100/r_j$ every time the stub~$P$
is charged to. Hence, the total cost charged to~$e$ is $O(\eps^{-2} \log (1/\eps)$.
The lemma now follows from the fact that the stub set has $O(n)$ edges
in total, by \autoref{cor:stub_set_linear}.
\end{proof}

\subparagraph{Completing the bound.}
Every relevant pair is of Type~1, Type~2, or Type~3. 
Hence, combining Lemmas~\ref{lem:Type_1},~\ref{lem:Type_2} and~\ref{lem:Type_3} implies
$\sum_{i=6}^L m_i = O(n\eps^{-2}\log(1/\eps))$.
As we already observed, $\sum_{i=0}^{5}m_i = O(n\eps^{-2})$,
and so $\sum_{i=0}^{L}m_i = O(n\eps^{-2}\log (1/\eps))$.
Together with Fact~\ref{fact:hpr}, this finishes the proof of \autoref{thm:main_introduction}(i).

\section{The lower bound} \label{sec:LB}
In this section we will prove part~(ii) of \autoref{thm:main_introduction}.
The proof combines two ingredients: 
a lemma that counts the number of well-separated pairs required for point sets 
with geometrically increasing pairwise distances,
and a \emph{hierarchy of antennas} that contains the point sets needed to apply the lemma.
In the following, it will be convenient to set $s := 1/\eps$
and work with an $s$-WSPD. 

\subsection{Counting pairs across distance scales} \label{ssec:counting_pairs}
We say that a point set is \emph{$\delta$-separated} if every two distinct points 
in it are at distance at least $\delta$ from each other.
\begin{lemma}\label{lem:certificate}
Let $\Psi$ be an $s$-WSPD of a metric space $(V,\dist)$, for some $s\geq 2$.
Let $\delta_j := 2^j$, and let $M_1,\ldots,M_L$ be subsets of $V$ 
such that each $M_j$ is $\delta_j$-separated.
Suppose that, for each $1\leq j\leq L$, we have a set $F_j\subseteq\binom{M_j}{2}$ of pairs 
such that $(s/100)\delta_j \leq \dist(p,q) < s \delta_j$ for every $\{p,q\}\in F_j$.
Then $\Psi$ contains $\Omega \left(|\cup_j F_j|\right)$ well-separated pairs. 
\end{lemma}
\begin{proof}
Define $F := \cup_j F_j$.
By the definition of WSPD, for every pair $\{p,q\}$ of points in~$V$---and,
hence, also for every pair in~$F$---there is a well-separated pair~$(A,B)\in\Psi$
such that $p\in A$ and $q\in B$ (or vice versa).
We assign each $\{p,q\} \in F$ to an arbitrary such pair.  
We claim that any pair $(A,B)\in \Psi$ can be assigned only constantly many pairs from~$F$,
which immediately implies the lemma.

To prove the claim, first consider a fixed set $F_j$ and
let $\{p,q\} \in F_j$ be assigned to $(A,B)$.  Since $\dist(p,q) < s\delta_j$
and $(A,B)$ is $s$-separated, we know that $\diam(A) < \delta_j$ and $\diam(B) < \delta_j$.
Hence, $A$ and $B$ can each contain at most one point from $M_j$, which
implies that $(A,B)$ is assigned at most one pair $\{p,q\} \in F_j$.

It remains to argue that $(A,B)$ can be assigned pairs $\{p,q\}$ from 
only constantly many sets~$F_j$. To see this, recall that 
$(A,B)$ being $s$-separated implies that
the distances between the points in~$A$ and the points in~$B$ 
are the same up to a constant factor, namely~$1+\tfrac{1}{s}$. 
Because $\dist(p,q) =\Theta(s\delta_j)$ for all 
pairs $\{p,q\}\in F_j$ and $\delta_j = 2^j$,
this means that $(A,B)$ can indeed only be assigned 
pairs from constantly many sets~$F_j$.
This finishes the proof of the lemma.
\end{proof}

\subsection{The hierarchical antenna and its analysis}
Let $Q$ be a parameter such that $Q=2^L$ for some integer~$L\geq 1$.
We now define the \emph{hierarchical antenna}~$T(Q)$ which, for a suitable choice
of $Q$, will form the basis of our lower-bound construction. 
The hierarchical antenna is tree that is an induced subgraph of the integer grid, as defined next.
\medskip

\begin{figure}[tbh]
    \centering
    \includegraphics[width=\textwidth]{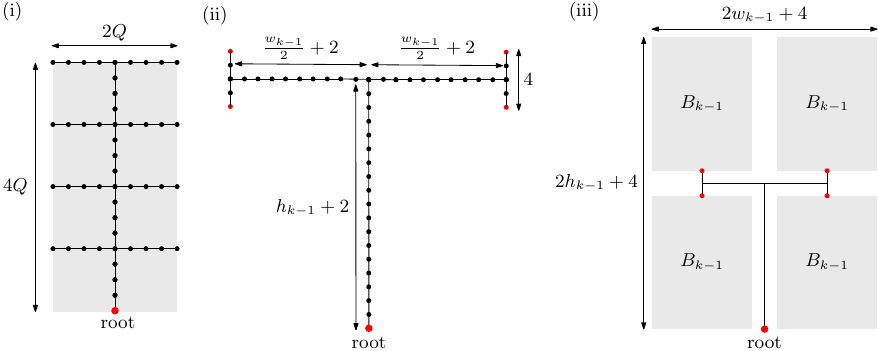}
    \caption{(i) A leaf block $B_0$. (ii) The connector gadget~$C_k$. Here $w_{k-1}$
                 and $h_{k-1}$ are the width and height of the block~$B_{k-1}$;
                 their values are given in \autoref{lem:grid}.
            (iii)~Constructing block $B_k$ from four blocks $B_{k-1}$
                  and a connector gadget~$C_k$}
    \label{fig:gadget-new}
\end{figure}

Define a level-0 block (or: \emph{leaf block})~$B_0$ to be an ``antenna''  with a backbone of 
$4Q$ edges and exactly $2Q$ ``teeth'', each of which is a path of $Q$~edges; see \autoref{fig:gadget-new}(i).
(Here and in the sequel, the term \emph{edge} refers to an edge of the integer grid.)
We call one endpoint of the backbone the \emph{root} of~$B_0$, and we
attach pairs of teeth to the backbone at distances $4,8,\ldots,4Q$ 
from the root.

A level-$k$ block $B_k$ consists of four level-$(k-1)$ blocks, two of which
are rotated by 180~degrees, that are connected to each other by a \emph{connector gadget}~$C_k$. 
We call these four blocks the \emph{children} of the level-$k$ block. The connector gadget,
which is shown in \autoref{fig:gadget-new}(ii), is a T-shape of which the
lower endpoint of the vertical bar will become the root of~$B_k$.
The horizontal bar of the T has two short vertical bars at its tips.
The four endpoints of these short vertical bars, shown in red in the figure,
are used to attach the four copies of~$B_{k-1}$. More precisely, these four
points are identified with the roots of these level-$(k-1)$ blocks.

We define our hierarchical antenna~$T(Q)$ to be the block~$B_L$. Observe that since $B_L$ is
an induced subgraph of the integer grid, it can be represented as a unit-disk graph.
The following observation follows immediately by induction
and the fact that $Q=2^L$. 
\begin{lemma} \label{lem:grid} $\mbox{}$

\begin{enumerate}[(i)]
\item The bounding box of $B_k$ has width $w_k := 2^k(2Q+4)+4$ 
      and height $h_k := 2^k(4Q+4)+4$.
\item $B_k$ has $4^k$ leaf blocks and $\Theta(4^k Q^2)$ vertices in total.
\item The distance of any vertex in $B_k$ to the root of $B_k$ is at most~$2^k(6Q+10)$.

\item The hierarchical antenna~$T(Q)$ has $Q^2/4^k$ level-$k$ blocks and $\Theta(Q^4)$ vertices in total.
\end{enumerate}
\end{lemma}

Our goal is to use \autoref{lem:certificate} and \autoref{lem:grid} to lower bound the size of an $s$-WPSD 
for the hierarchical antenna~$T(Q)$. To be able to apply \autoref{lem:certificate}, we need to define 
suitable sets~$M_j$ and $F_j$. This is done as follows.
\medskip

We start by defining the sets~$M_j$.
For $j=0,1,\ldots,L$, define $\delta_j := 2^j$ and recall that $Q = 2^L$.
Let $\B_0$ be the collection of all leaf blocks of $T(Q)$; each of them is a copy of the antenna~$B_0$.
For each leaf block $B\in\B_0$, select a set $V_j(B)$ of vertices as follows:
for each of the $2Q$ teeth of~$B$, select $Q/\delta_j$ equally-spaced vertices,
at distances $\delta_j, 2\delta_j, \ldots,  Q$ from the backbone.
Let $M_j := \bigcup_{B\in\B_0} V_j(B)$ be the union of the selected vertices over all teeth of all the leaf blocks. 
Clearly, any two selected vertices on the same tooth are at least $\delta_j$ apart.
Moreover, any two selected vertices in different teeth (of the same block or of
different blocks) are more than $\delta_j$ apart, because the unique shortest
path in $T(Q)$ must be going via the backbone(s) of the teeth, which are at least distance~$\delta_j$
from the selected vertices. Hence, we have the following observation.
\begin{observation} \label{obs:Mj}
Each set $M_j$ is $\delta_j$-separated.
\end{observation}
We now define the sets~$F_j\subseteq\binom{M_j}{2}$. To simplify the exposition,
it will be convenient to set $F_0 := \emptyset$ and define non-empty
$F_j$'s only for $1\leq j\leq L$, as follows.

Let $\B_j$ be the collection of all level-$j$ blocks in~$T(Q)$. Any block~$B\in \B_j$
has four children, which are level-$(j-1)$ blocks, that are connected to each other using the connector gadget~$C_j$.
Let $B_{\mathrm{ne}}$ and $B_{\mathrm{nw}}$ be the two children of~$B$ whose roots are
at distance~$w_{j-1}+8$ from each other. (The two ``topmost'' blocks in \autoref{fig:gadget-new}(iii)
satisfy this condition, for example.) 
Then we define 
\[
F_j(B) := \{ \{u,v\} : u\in B_{\mathrm{ne}}\cap M_j \mbox{ and } v\in B_{\mathrm{nw}}\cap M_j \}
\]
and we set $F_j := \bigcup_{B\in\B_j} F_j(B)$.
\begin{lemma} \label{lem:Fj}
For any $1\leq j\leq L$, we have $2^j Q \leq \dist(u,v) < 100\cdot 2^j Q$ for all $\{u,v\}\in F_j$,
and we have $|F_j| = \Theta(Q^6)$. Moreover, the sets $F_j$ are pairwise disjoint.
\end{lemma}
\begin{proof}
Consider a pair $\{u,v\}\in F_j(B)$ for some block $B\in \B_j$. Then $\dist(u,v)$ is larger
than the distance between the roots of $B_{\mathrm{ne}}$ and $B_{\mathrm{nw}}$, and so
$\dist(u,v) > w_{j-1} + 8 > 2^j Q$.
On the other hand, $\dist(u,v)$ cannot be more than the distance between these roots plus
the distances from the roots to $u$ and $v$, respectively. Using \autoref{lem:grid}
we conclude that 
\[
\begin{array}{lll}
\dist(u,v) & \leq & w_{j-1} + 8 + 2\cdot  2^j(6Q+10)  \\
            & < & \left(2^j(2Q+4)+4\right) + 8 +  2^{j+1}(6Q+10)  \\
            & < & 100 \cdot 2^j Q.
\end{array}
\]
We conclude that $2^j Q \leq \dist(u,v) < 100\cdot 2^j Q$, as required.

Next, we prove that $|F_j| = \Theta(Q^6)$
By \autoref{lem:grid}, the children $B_{\mathrm{ne}}$ and $B_{\mathrm{nw}}$ of a block~$B\in \B_j$
have $4^{j-1}$ leaf blocks, and $M_j$ selects $2Q^2/\delta_j$ vertices from each of them. Hence,
\[
|F_j| = \sum_{B\in \B_j} |F_j(B)| 
      = \sum_{B\in \B_j} \left( 4^{j-1} \cdot 2Q^2/\delta_j \right)^2
      = \left( Q^2/4^j \right) \cdot \left( 4^{j-1} \cdot 2Q^2/\delta_j \right)^2
      = \Theta(Q^6),
\]
where the last step uses that~$\delta_j=2^j$.

Finally, to prove that the sets $F_j$ are pairwise disjoint, we
observe that $\{u,v\}\in F_j$ implies that $u$ and $v$ are in the same level-$j$
block~$B$ but in different level-$(j-1)$ blocks. Because of the tree structure
of $T(Q)$, this can happen for only one value of~$j$.
\end{proof}
We can use \autoref{lem:certificate} to lower bound the size of an $s$-WSPD of~$T(Q)$
when $s := 100Q$. 
\begin{corollary}\label{cor:gadget-new}
Let $s := 100 Q$. Then any $s$-WSPD $\Psi$ of $T(Q)$ contains 
$\Omega (Q^6 \cdot \log Q)$ pairs. 
\end{corollary} 
\begin{proof}
\autoref{obs:Mj} implies that the sets $M_j$ defined above satisfy the conditions
of \autoref{lem:certificate}. Because $s := 100Q$, \autoref{lem:Fj} implies that 
the sets $F_j$ satisfy those conditions. Since the
sets $F_j$ are pairwise disjoint, we thus obtain
\[
|\Psi| \geq \Omega\Bigl(\sum_{j=1}^L |F_j|\Bigr) = \Omega(L\cdot Q^6) = \Omega (Q^6 \log Q).
\]
\end{proof}

\subsection{Proof of part~(ii) of \autoref{thm:main_introduction}}

\begin{figure}
    \centering
    \includegraphics{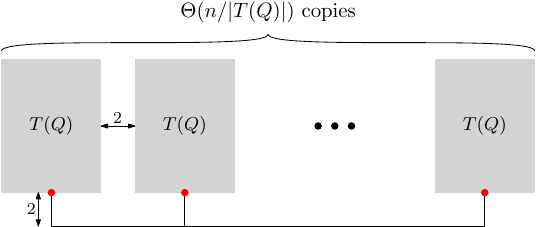}
    \caption{The full lower bound construction. The small vertical segments incident to
    the roots of the trees~$T(Q)$ are paths of length~2.}
    \label{fig:global-construction}
\end{figure}

Recall that the size of $T(Q)$ is $\Theta(Q^4)$. Hence, \autoref{cor:gadget-new} 
gives us a tree $T(Q)$ of size $\Theta(s^4)$ that requires an $s$-WSPD of size $\Omega(s^4 \cdot s^2 \log s)$.
In other words, $T(Q)$ requires a $(1/\eps)$-WSPD of size $\Omega(|T(Q)| \cdot (1/\eps)^2 \log (1/\eps))$,
where $|T(Q)|=\Theta(1/\eps^4)$. To prove part~(ii) of \autoref{thm:main_introduction},
we simply place $\Theta(n/|T(Q)|)$ copies of $T(Q)$ next to each other and we connect their
roots as shown in \autoref{fig:global-construction}. Note that the tree $T^*$ that we obtain
in this manner is still an induced subgraph of the integer grid.

The total number of vertices in $T^*$ is $\Theta(n)$. 
To finish the proof of the lower bound, we need to argue that the total number 
of pairs in an $s$-WSPD of $T^*$ is the sum of the required sizes for the
$s$-WSPDs we would need for each of copies of~$T(Q)$. To see this, we can
use \autoref{lem:certificate} again: simply define the sets $M_j$ and $F_j$
for the tree~$T^*$ to be the unions of the corresponding sets for the copies of~$T(Q)$.
Clearly, the global sets $M_j$ and $F_j$ still satisfy the requirements
needed to apply \autoref{lem:certificate}. This finishes the proof of the lower bound.

\section{Construction algorithm} \label{sec:construction}

We have shown matching upper and lower bounds on the size of a WSPD for unit-disk graphs (Sections~\ref{sec:upper-bound} and~\ref{sec:LB}).  In this section we give a deterministic algorithm that, given a set $V$ of $n$ points in the plane, constructs such a WSPD in
$O(n\log n + n\eps^{-2} \log(1/\eps))$ time; its output has the size guaranteed by the upper bound.

The algorithm produces two kinds of pairs. \emph{Short pairs} consist of sets of Euclidean diameter at most one; such a set induces a clique in $G$ and these pairs are obtained by filtering the pairs obtained from a Euclidean WSPD (Step~1). \emph{Long pairs} are produced by a Callahan--Kosaraju style recursion~\cite{callahan-kosaraju} on a hierarchy of weak nets, in which one cluster is split at a time and separation is tested in constant time with the help of sparse covers (Step~2). Both the nets and the covers are built on a plane $3$-spanner $H$ of $G$; since a WSPD
only needs approximate distances up to constant factors, this suffices. 

\begin{figure}[tbh]
    \centering
    \includegraphics[scale=0.95]{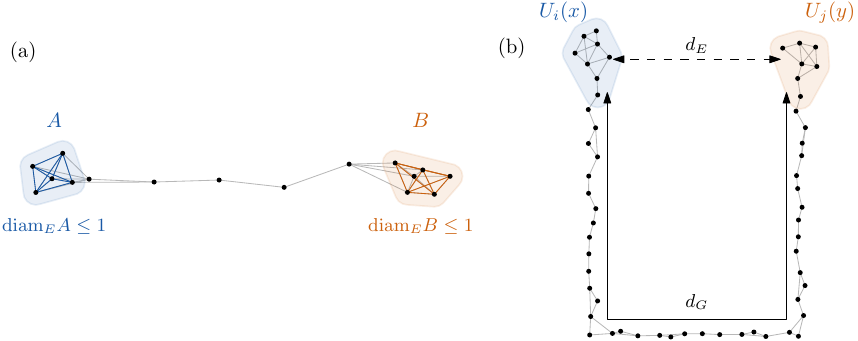}
\caption{(a) A short pair: each set has Euclidean diameter at most one, so it
  induces a clique in $G$. (b) A long pair whose clusters are close in the plane but
  far apart in $G$; such pairs must be separated with respect to graph distance.}
    \label{fig:short-long}
\end{figure}

\subsection{Preliminaries}\label{sec:prelim}
Let $H$ be the subgraph of the Delaunay triangulation of $V$ consisting of the Delaunay edges of length at most one. Then $H$ is a plane graph with $O(n)$ edges, it can be computed in $O(n \log n)$ time, and Chan and Skrepetos~\cite{chan-skrepetos} showed that it is a $3$-spanner of $G$, that is, for every pair of vertices $u, v$ in $V$ we have 
\begin{equation} \label{eq:spanner}
\dG(u,v) \le \dH(u,v) \le 3 \dG(u,v).
\end{equation} 
In particular $H$ is connected, and every edge of $H$ has length at most one. 

\begin{figure}[tbh]
    \centering
    \includegraphics[scale=0.8]{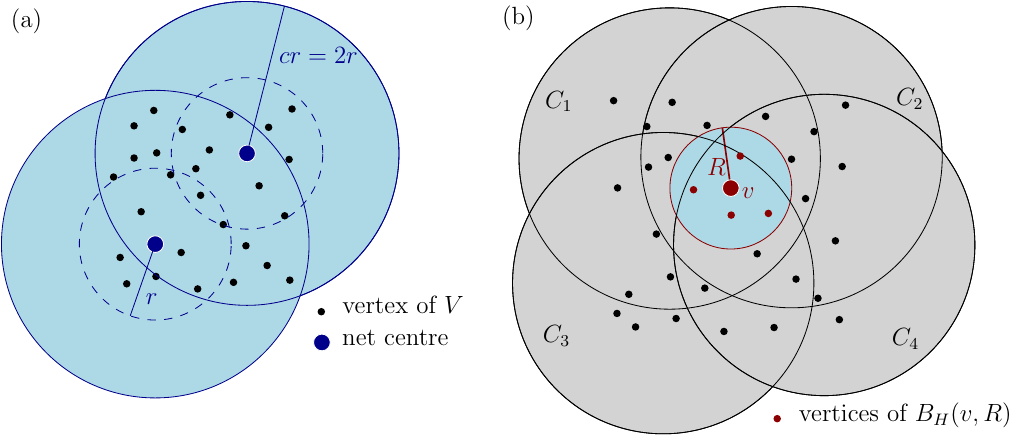}
  \caption{(a) A weak $r$-net with covering factor $c=2$. (b) A sparse cover at radius
    $R$ with constant $c=5$; the highlighted points form $B_H(v,R)$, which is contained in
    the cluster $C_2$. For simplicity, the figure uses Euclidean distance in place of
    $\dH$.}
    \label{fig:nets-covers}
\end{figure}

Our algorithm heavily relies on weak nets and sparse covers which we define next. Both structures are illustrated in figure~\ref{fig:nets-covers}.

\begin{definition}[Weak net]\label{def:weak-net}
  Let $T\subseteq V$, $r>0$ and $c\ge1$. A \emph{weak $r$-net of $T$ with covering factor $c$} is a set $N\subseteq T$ such that (i)~$\dH(x,y)\ge r$ for all distinct $x,y\in N$, and (ii)~for every $u\in T$ there is an $x\in N$ with $\dH(u,x)\le cr$.
\end{definition}

For $v\in V$ and $R>0$, let $B_H(v,R)=\{w\in V | \dH(v,w)\le R\}$, i.e. the set of vertices in $V$ within graph distance $R$ from $v$ in $H$.

\begin{definition}[Sparse cover]\label{def:sparse-cover}
  Let $R>0$ and $c \ge 1$. A \emph{sparse cover at radius $R$ with constant $c$} is a set $\mathcal C$ of clusters of $V$ such that
  (i)~for every $v\in V$ some cluster $C \in \mathcal C$ contains $B_H(v,R)$;
  (ii)~for every $C\in\mathcal C$ $\diam_H C \le cR$; and
  (iii)~every vertex belongs to at most $c$ clusters.
\end{definition}

Le and Wulff-Nilsen~\cite[Theorem~5 and Lemma~1]{LW21} showed that both weak nets and sparse covers can be computed in linear time for planar graphs.

\begin{fact}[Le and Wulff-Nilsen~\cite{LW21}]\label{fact:LW}
  There is a constant $c \ge 1$ such that, for every $n$-vertex edge-weighted planar graph $H$, the following can be computed in $O(n)$ time:
  \begin{description}
    \item[$(a)$] given $T\subseteq V$ and $r>0$, a weak $r$-net $N$ of $T$ with covering factor $c$, together with an assignment of every $u \in T$ to some $x\in N$ with $\dH(u,x)\le cr$;
    \item[$(b)$] given $R>0$, a sparse cover at radius $R$ with constant $c$.
  \end{description}
\end{fact}

From now on, $c$ denotes the constant of Fact~\ref{fact:LW}. 

Set $r_i = 36 \cdot 2^i$, $R_i=\frac{24c\,r_i}{\eps}$, $s=\frac{64c^2 r_0}{\eps}$ and $L = \min\{i \ge 0 | r_i \ge n\}$. Since $H$ is connected and every edge has length at most one, its diameter is at most $n-1 < r_L$. Hence, $L\leq \lceil\log n\rceil$.

\subsection{Step 1: short pairs}\label{sec:step1}
Compute a Euclidean WSPD of $V$ in which every pair $(A,B)$ satisfies
\begin{equation}\label{eq:euclid-ws}
  \max\{\diam_E A,\ \diam_E B\} \le \frac{\dE(A,B)}{s}.
\end{equation}
Any Euclidean WSPD of $V$ with the following properties can be used: every pair of points of $V$ lies in exactly one of its pairs, every pair satisfies~\eqref{eq:euclid-ws}, its sets are nodes of a rooted tree with $O(s^2n)$ nodes, and it has $O(s^2n)=O(n\eps^{-2})$ pairs and can be computed in $O(n\log n+s^2n)=O(n\log n+n\eps^{-2})$ time. Examples are the WSPD of Callahan and Kosaraju~\cite{callahan-kosaraju} with separation constant $2s$, whose sets are nodes of a fair split tree and lie in balls of a common radius $\rho$ at distance at least $2s\rho$, and the quadtree-based WSPD of Gudmundsson and Wong~\cite{DBLP:conf/soda/GudmundssonW26} with $1/s$ in place of $\eps$. We compute the bounding box of every tree node bottom-up. We keep a well-separated pair $(A,B)$ if and only if the bounding-box diagonals of $A$ and of $B$ both have length at most one. These retained pairs are the \emph{short pairs}.

\subsection{Step 2: long pairs}\label{sec:step2}

\begin{figure}[t]
  \centering
\includegraphics[width=\textwidth]{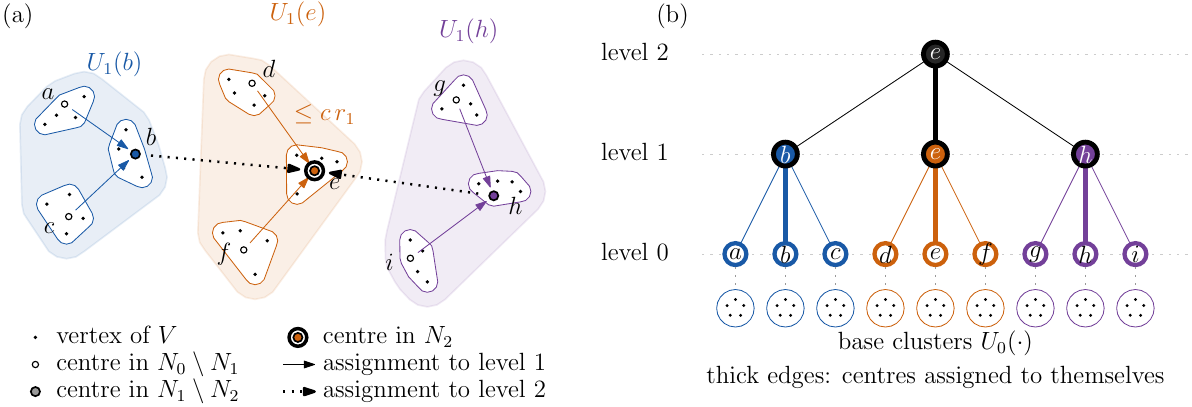}
  \caption{The cluster hierarchy with three levels. (a)~Base clusters $U_0(\cdot)$
    (white) are grouped into the level-$1$ clusters $U_1(b)$, $U_1(e)$ and $U_1(h)$
    (shaded); arrows show the assignments, each of length at most $c\,r_i$ in $\dH$.
    For simplicity, Euclidean distance is used in place of $\dH$. (b)~The
    corresponding tree $\mathcal T$, whose node $(i,x)$ represents $U_i(x)$. 
    }
  \label{fig:hierarchy}
\end{figure}

\paragraph*{The cluster hierarchy}
Using Fact~\ref{fact:LW}(a) on $H$, compute a weak $r_0$-net $N_0$ of $T=V$ and, for $i=1,\dots,L$, a weak $r_i$-net $N_i$ of $T=N_{i-1}$, together with the assignments. We then reassign every centre $x\in N_i$ to itself, which is allowed since $\dH(x,x)=0\le c\,r_i$. Thus $V \supseteq N_0 \supseteq N_1 \supseteq \cdots \supseteq N_L$. The assignments define a tree $\mathcal T$ whose nodes are the pairs $(i,x)$ with $0 \le i \le L$ and $x \in N_i$; the parent of $(i,x)$, for $i < L$, is $(i+1,z)$, where $z$ is the centre to which $x$ is assigned in $N_{i+1}$. For a node $(i,x)$ let $U_i(x)$ be the set of vertices $u\in V$ whose assigned level-$0$ centre is a level-$0$ descendant of $(i,x)$. We call $x$ the \emph{centre} and $i$ the \emph{level} of the cluster $U_i(x)$; the clusters $U_0(x)$ are the \emph{base clusters}. See figure~\ref{fig:hierarchy} for an illustration.

\begin{lemma}\label{lem:hierarchy}
  The hierarchy has the following properties.
  \begin{description}
    \item[$(i)$] For every $i$, the sets $U_i(x)$, $x\in N_i$, partition $V$; moreover $x\in U_i(x)$, and the clusters of the children of $(i,x)$ partition $U_i(x)$.
    \item[$(ii)$] If $u\in U_i(x)$ then $\dG(u,x) \le \dH(u,x)<2c r_i$. Hence $\diam_G U_i(x)<4c r_i$.
    \item[$(iii)$] Distinct $x,y\in N_i$ satisfy $\dH(x,y)\ge r_i$ and $\dG(x,y)\ge r_i/3$.
    \item[$(iv)$] $|N_i|\le 1+2n/r_i$ for every $i$, and $|N_L|=1$. So, $\mathcal T$ is a tree and $\sum_{i=0}^{L} |N_i|=O(n)$.
  \end{description}
\end{lemma}

\begin{proof}
  $(i)$ Every vertex has exactly one assigned level-$0$ centre, and every node has exactly one ancestor at each higher level, so the sets $U_i(x)$ partition $V$ and the children of $(i,x)$ partition $U_i(x)$. Since every centre is assigned to itself, $(0,x), (1,x), \dots , (i,x)$ is a path in $\mathcal T$ and $x$ is assigned to itself at level $0$; hence $x\in U_i(x)$.

  $(ii)$ Let $u\in U_i(x)$, and let $a_0, a_1, \dots , a_i = x$ be the centres of the ancestors of $u$'s base cluster, so that $u$ is assigned to $a_0 \in N_0$ and $a_{j-1}$ is assigned to $a_j\in N_j$. By Fact~\ref{fact:LW}(a) and the triangle inequality,
  $$\dH(u,x) \le \sum_{j=0}^{i} c r_j = c r_0 (2^{i+1}-1) < 2cr_i.$$
  The bound for $\dG$ follows from~\eqref{eq:spanner}, and the diameter bound from the triangle inequality through $x$.

  $(iii)$ The first inequality follows from Definition~\ref{def:weak-net}(i) and the second follows from~\eqref{eq:spanner}.

  $(iv)$ If $|N_i| \ge 2$, consider $x \in N_i$ and a shortest path in $H$ from $x$ to another centre of $N_i$, which has length at least $r_i$ by~(iii). Along this path the distance from $x$ starts at $0$ and grows by at most one per edge, so the path has at least $\lceil r_i/2\rceil$ vertices at $H$-distance less than $r_i/2$ from $x$. By~(iii) and the triangle inequality, the open balls of radius $r_i/2$ about distinct centres are disjoint. Hence $|N_i|\cdot r_i/2\le n$. Since $H$ is connected with edges of length at most one, its diameter is at most $n-1 < r_L$, so by~(iii) $|N_L| = 1$. Finally, $\sum_i |N_i| \le (L+1) + 2n\sum_{i \ge 0} (1/r_i) = O(n)$.
\end{proof}

\paragraph*{The separation test}
For every level $i$, use Fact~\ref{fact:LW}(b) to compute a sparse cover $\mathcal C_i$ at radius $R_i$ with constant $c$. For a vertex $u$ let $\Lambda_i(u)=\{C \in \mathcal C_i: u\in C\}$; by Definition~\ref{def:sparse-cover}(iii), $|\Lambda_i(u)| \le c$, so two such lists, stored as sorted lists of cluster identifiers, can be compared in constant time. We store $\Lambda_0(x)$ for every $x \in N_0$ and, for $i \ge 1$, $\Lambda_i(x)$ for every $x\in N_{i-1}$.

\begin{lemma}\label{lem:test}
  For every $i$ and all $u,v\in V$,
  $$\Lambda_i(u)\cap\Lambda_i(v)=\emptyset\;\Longrightarrow\;\dG(u,v)>R_i/3,
    \quad \mathrm{and } \quad 
    \Lambda_i(u)\cap\Lambda_i(v)\neq\emptyset\;\Longrightarrow\;\dG(u,v)\le cR_i.$$
\end{lemma}
\begin{proof}
  If $\dH(u,v)\le R_i$, then a cluster containing $B_H(u,R_i)$ contains both $u$ and $v$, so the lists intersect; hence disjoint lists imply $\dH(u,v)>R_i$, and $\dG(u,v)>R_i/3$ since $H$ is a $3$-spanner of $G$. If the lists share a cluster $C$, then $\dG(u,v) \le \dH(u,v) \le \diam_H C \le cR_i$.
\end{proof}

\paragraph*{The recursion}
The long pairs are produced by a recursive procedure, $\textsc{Process}(i,x,j,y)$, that receives two clusters $U_i(x)$ and $U_j(y)$ and outputs well-separated pairs covering all pairs of vertices in $U_i(x)\times U_j(y)$, except possibly some that are close in $G$ and will be covered by the short pairs. It is first called on every pair of sibling clusters: for each level $i=0, \dots, L-1$, each $z \in N_{i+1}$, and each unordered pair of distinct children $(i,x)$ and $(i,y)$ of $(i+1,z)$, we call $\textsc{Process}(i,x,i,y)$. These are the \emph{initial calls}.

A call $\textsc{Process}(i,x,j,y)$ proceeds as follows. Let $\ell=\max\{i,j\}$ be the larger of the two levels. First, it tests whether the lists $\Lambda_\ell(x)$ and $\Lambda_\ell(y)$ are disjoint. If they are, the call outputs the pair $(U_i(x),U_j(y))$ and stops. Otherwise, if $\ell=0$, the call stops without output; we show in Section~\ref{sec:correctness} that the vertex pairs of such a call are covered by the short pairs. In the remaining case, the call splits the cluster of larger level and recurses on its children. 

\begin{figure}[t]
  \centering
  \includegraphics[width=\textwidth]{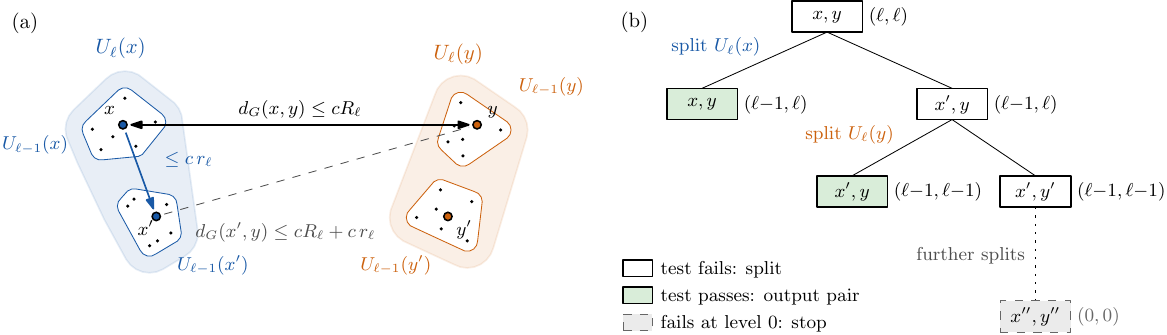}
 \caption{(a) A failed call on $U_\ell(x)$ and $U_\ell(y)$: the test gives
  $d_G(x,y)\le cR_\ell$, and splitting $U_\ell(x)$ replaces $x$ by a child centre $x'$
  with $d_H(x,x')\le c\,r_\ell$, so $d_G(x',y)\le cR_\ell+c\,r_\ell$ (not to scale).
  (b) One branch of the recursion, with the levels of each call in parentheses; the
  cluster of larger level is always split, so the levels differ by at most one.}
  \label{fig:recursion}
\end{figure}

The pairs output by \textsc{Process} are the \emph{long pairs}, and the final output consists of the short and the long pairs. Note that the cluster to be split is chosen by level alone. The stored
lists suffice for all tests, by the following invariant.

\begin{lemma}\label{lem:levels}
  Every call $\textsc{Process}(i,x,j,y)$ satisfies $|i-j| \le 1$ and $x \ne y$, and the clusters $U_i(x)$ and $U_j(y)$ are disjoint. Consequently, with $\ell = \max\{i,j\}$ and $k=\min\{i,j\}$, both centres lie in $N_k$, and $\Lambda_\ell(x)$ and $\Lambda_\ell(y)$ are stored.
\end{lemma}

\begin{proof}
  Initial calls have equal levels. Splitting at levels $(\ell,\ell)$ produces levels $(\ell-1,\ell)$ up to order, and splitting the larger level of $(\ell-1,\ell)$ produces $(\ell-1,\ell-1)$; so $|i-j| \le 1$ is preserved. Initial clusters are disjoint by Lemma~\ref{lem:hierarchy}(i), and splitting replaces a cluster by a subset, so disjointness is preserved; as $x \in U_i(x)$ and $y \in U_j(y)$, we get $x \ne y$. Since the centre sets are nested and $k \ge \ell-1$, both centres lie in $N_k$, which is $N_{\ell-1}$ if $\ell \ge 1$ and $N_0$ if $\ell=0$; their lists at level $\ell$ are therefore stored.
\end{proof}

\subsection{Correctness}\label{sec:correctness}

\begin{lemma}\label{lem:separated}
  Every output pair is $(1/\eps)$-well-separated in $G$.
\end{lemma}

\begin{proof}
  \textit{Long pairs.} Let $(U_i(x),U_j(y))$ be output and $\ell=\max\{i,j\}$. The sets are nonempty and disjoint by Lemmas~\ref{lem:hierarchy} and~\ref{lem:levels}. By Lemma~\ref{lem:test}, $\dG(x,y) > R_\ell/3 = 8cr_\ell/\eps$. By Lemma~\ref{lem:hierarchy}(ii), every vertex of either cluster is within graph distance $2cr_\ell$ of its centre, so
  $$ \dG(U_i(x),U_j(y)) > \frac{8c r_\ell}{\eps}-4c r_\ell \ge \frac{4c r_\ell}{\eps},$$
  using $\eps\le 1$. Both clusters have graph diameter less than $4c r_\ell$, so the pair is $(1/\eps)$-well-separated.

  \textit{Short pairs.} Let $(A,B)$ be a pair produced in Step~1. Each of $A,B$ has bounding-box diagonal, and hence Euclidean diameter, at most one, so it induces a clique in $G$ and $\diam_G A=\diam_E A$, $\diam_G B=\diam_E B$. Since $\dG \ge \dE$, inequality~\eqref{eq:euclid-ws} and $s \ge 1/\eps$ give $\max\{\diam_G A,\diam_G B\} \le \dE(A,B)/s \le \eps \dG(A,B)$.
\end{proof}

\begin{lemma}\label{lem:cover}
  For all pairs of vertices $u,v\in V$, some output pair $(A,B)$ has $u\in A$ and $v\in B$.
\end{lemma}
\begin{proof}
  We first show that $u,v$ are covered by a long pair or $\dG(u,v)\le 25c^2r_0/\eps$. Suppose first that $u$ and $v$ lie in different base clusters. By Lemma~\ref{lem:hierarchy}(iv) $\mathcal T$ is a tree, so the base clusters of $u$ and $v$ have a lowest common ancestor $(i+1,z)$, and $u$ and $v$ lie in the clusters of two distinct children $(i,x)$, $(i,y)$ of it. Hence, exactly one initial
  call contains $(u,v)$ in the product of its two clusters. A split replaces the product of a call by a partition of it into the products of its subcalls (Lemma~\ref{lem:hierarchy}(i)), so exactly one call at each depth of the recursion contains $(u,v)$, and this branch is finite because the sum of the two levels decreases with every split. The branch ends either by outputting a pair, which
  covers $u,v$, or with a failed test at $\ell=0$. In the latter case the call is $\textsc{Process}(0,x'',0,y'')$ with $u\in U_0(x'')$, $v\in U_0(y'')$, and $\dG(x'',y'')\le cR_0$ by Lemma~\ref{lem:test}. By Fact~\ref{fact:LW}(a), $u$ and $v$ lie within $H$-distance $cr_0$ of $x''$ and $y''$ respectively, so
  $$\dG(u,v) \le 2c r_0 + cR_0 = 2c r_0+\frac{24c^2r_0}{\eps} \le \frac{25c^2r_0}{\eps},$$
  where the last step uses $c \ge 1$ and $\eps \le 1/2$. If $u$ and $v$ lie in the same base cluster $U_0(x'')$, then $\dG(u,v)\le 2c\,r_0$, and the same bound holds.

  In the second case, let $(A,B)$ be the pair of the Euclidean WSPD of Step~1 with $u \in A$, $v \in B$. By~\eqref{eq:euclid-ws} and our choice of parameter values,
  $$\max\{\diam_E A,\diam_E B\} \le \frac{\eps}{64c^2 r_0} \dE(A,B) \le \frac{\eps}{64c^2r_0} \|u-v\| \le \frac{\eps}{64c^2r_0} \dG(u,v) \le \frac{25}{64}.$$

  The bounding-box diagonal of a planar point set is at most $\sqrt2$ times its diameter, so both diagonals are at most $25\sqrt2/64<1$, and $(A,B)$ was retained as a short pair.
\end{proof}
By Lemmas~\ref{lem:separated} and~\ref{lem:cover}, the output is a $(1/\eps)$-WSPD of $G$.

\subsection{Size and running time}\label{sec:analysis}

The analysis hinges on the number $K$ of calls to \textsc{Process}, which bounds both the number of long pairs and the time spent in Step~2.

\begin{lemma}\label{lem:call-distance}
  Every call $\textsc{Process}(i,x,j,y)$ with $k=\min\{i,j\}$ satisfies $\dG(x,y)\le 49c^2r_k/\eps$.
\end{lemma}
\begin{proof}
  If it is an initial call then $x$ and $y$ are assigned to a common centre $z \in N_{k+1}$, so $\dG(x,y) \le \dH(x,y) \le 2c r_{k+1} = 4c r_k$. Otherwise, by the proof of Lemma~\ref{lem:levels}, the parent call has larger level $k+1$, its test failed, and so its centres are at graph distance at most $cR_{k+1}$ by Lemma~\ref{lem:test}. The split replaces one of these centres by the centre of one of its children, which is assigned to it and therefore at distance at most $cr_{k+1}$. Hence
  $$\dG(x,y) \le cr_{k+1} + cR_{k+1} = 2c r_k + \frac{48c^2r_k}{\eps} \le \frac{49c^2r_k}{\eps},$$  
  using $c \ge 1$ and $\eps \le 1/2$. Since $4c r_k \le 49c^2r_k/\eps$, the bound holds in both cases.
\end{proof}

\begin{lemma}\label{lem:calls}
  $K=O(n\eps^{-2}\log(1/\eps))$.
\end{lemma}
\begin{proof}
  \emph{Each unordered pair of nodes of $\mathcal T$ is processed at most once.} The products of the initial calls are pairwise disjoint, and every split partitions the product of its call among its subcalls, so two calls in different branches have disjoint products and hence distinct node pairs (the products are nonempty by Lemma~\ref{lem:hierarchy}(i)). Within one branch, the sum of the two levels strictly decreases, so no node pair recurs. This holds even for a node with a single child, whose cluster coincides with that of its child: the two nodes differ in level.

  \emph{Charging.} Charge the call $\textsc{Process}(i,x,j,y)$ to the triple $(k,\{x,y\})$ with $k=\min\{i,j\}$. Since a node is determined by its level and centre, and $|i-j| \le 1$, the node pairs that receive the charge $(k,\{x,y\})$ are among $\{(k,x),(k,y)\}$, $\{(k,x),(k+1,y)\}$ and $\{(k+1,x),(k,y)\}$. Hence, at most three calls receive the same charge, and $K$ is at most three times the number of triples $(k,\{x,y\})$ with distinct $x,y\in N_k$  and $\dG(x,y) \le 49c^2r_k/\eps$; here $x,y\in N_k$ and $x \ne y$ hold by Lemma~\ref{lem:levels}, and the distance bound by Lemma~\ref{lem:call-distance}.

  \emph{Counting.} %
  By Lemma~\ref{lem:hierarchy}(iii) the set $N_k$ is $(r_k/3)$-separated in $G$, and $\frac{49c^2r_k}{\eps} \le \frac{64c^2r_k}{\eps}$. 
  The counting bounds of Section~\ref{sec:upper-bound} (\autoref{cor:stub_set_linear}, Lemmas~\ref{lem:Type_1},~\ref{lem:Type_2} and~\ref{lem:Type_3}) apply to any sequence of $(r_k/3)$-separated centre sets at level $k$; their proofs use only centre separation and the growth of the scales. Applied with parameter $\eps/(6c^2)$, they bound the number of such triples by $O(n (6c^2/\eps)^{2} \log (6c^2/\eps))$. Levels with fewer than two centres contribute nothing. Since $c$ is a constant, $K=O(n\eps^{-2} \log (1/\eps))$.
\end{proof}

\begin{proof}[Proof of running time in Theorem~\ref{thm:main_introduction}(i)] We go through each of the stated results separately. 
\begin{description}
    \item[Correctness.] The correctness follows immediately from Lemmas~\ref{lem:separated} and~\ref{lem:cover}.
    \item[Size.] Each call outputs at most one long pair, so there are at most $K=O(n\eps^{-2} \log(1/\eps))$ long pairs, according to Lemma~\ref{lem:calls}, and Step~1 contributes $O(n \eps^{-2})$ short pairs, hence, in total $O(n\eps^{-2} \log(1/\eps))$ well-separated pairs. 
    \item[Proprocessing.] The spanner $H$ takes $O(n \log n)$ time. For each of the $L+1=O(\log n)$ levels, Fact~\ref{fact:LW} gives the weak net with its assignment and the sparse cover in $O(n)$ time, and the lists $\Lambda_i$ are extracted in $O(n)$ time; this totals $O(n \log n)$. By Lemma~\ref{lem:hierarchy}(iv) the tree $\mathcal T$ has $O(n)$ nodes and the number of stored list entries is at most $c|N_0|+c\sum_{i=1}^{L} |N_{i-1}| = O(n)$. 
    \item[Steps 1 and 2.] Step~1 takes $O(n \log n+n \eps^{-2})$ time, since with the stored bounding boxes each pair is filtered in constant time. In Step~2, scanning $\mathcal T$ to enumerate the initial calls takes $O(n)$ time plus constant time per initial call. In every call, the separation test, the output of a pair of node pointers, and the enumeration of each subcall take constant time, and the latter is charged to the subcall. Step~2 therefore takes $O(n+K)$ time.

    Summing gives a total running time of $O(n\log n+n\eps^{-2}\log(1/\eps))$.  
\end{description}
\end{proof}

\section{From cover WSPD to partition WSPD}
\label{sec:partition}

So far, we have focused on the cover definition of the WSPD. The lower bound in Section~\ref{sec:LB} applies to both cover and partition WSPDs. In this section, we will adapt the construction of Section~\ref{sec:construction} so that every pair of distinct vertices is covered exactly once. 
This extends the proof of Theorem~\ref{thm:main_introduction}(i) from cover to partition WSPDs.
The main idea is to split each short pairs into a constant number of pairs, using the set of base clusters from the net-tree.

\begin{theorem}\label{thm:udg-partition}
Let $V$ be a set of $n$ points in the plane whose unit-disk graph $G$ is connected, and let $0<\eps\le 1/2$. A partition $(1/\eps)$-WSPD of $G$ of size $O(n\eps^{-2}\log(1/\eps))$ can be constructed in $O(n\log n+n\eps^{-2}\log(1/\eps))$ time.
\end{theorem}

\begin{proof}
 Let $x,y \in N_0$ be centres at level zero of the net-tree~$\mathcal T$.
The pair of base clusters $(U_0(x),U_0(y))$ is ``marked'' if no long pair contains $x$ and $y$ on opposite sides.
By Lemma~\ref{lem:cover}, every pair of distinct vertices in a marked
pair $(U_0(x),U_0(y))$ is covered by a short pair.

We keep all long pairs from \hyperref[sec:step2]{Step~2}. For each short pair $(A,B)$ obtained from the Euclidean WSPD in \hyperref[sec:step1]{Step~1}, and for each marked $(U_0(x),U_0(y))$, if $A\cap U_0(x) \neq \emptyset$ and $B\cap U_0(y) \neq \emptyset$, then we add~$\bigl(A\cap U_0(x),\,B\cap U_0(y)\bigr)$ to the partition WSPD.

Next, we prove correctness. By the construction in Section~\ref{sec:step2}, each vertex pair is
covered by at most one long pair. Whenever no long pair covers the two vertices, Lemma~\ref{lem:cover} shows that the unique Euclidean WSPD pair covering the two vertices was retained in \hyperref[sec:step1]{Step~1}. Our rule then outputs exactly one pair of parts for these vertices, namely the one selected by their base clusters. Well-separation follows from Lemma~\ref{lem:separated}, since taking subsets preserves well-separation.

We bound the number of pairs in our partition WSPD. If $(A,B)$ is a short pair, then $A$ has Euclidean diameter and therefore graph diameter $\leq 1$. Two base clusters intersecting~$A$ have centres at graph distance at most $2cr_0+1$, since the radius of a base cluster is $\leq cr_0$ by Fact~\ref{fact:LW}(a). However, base centres in $N_0$ are $(r_0/3)$-separated, where $r_0/3=12$, by Lemma~\ref{lem:hierarchy}(iii). Fact~\ref{fact:packing} therefore gives only $O(1)$ such centres, as $c$ and $r_0$ are constants. Since~$A$ only intersects $O(1)$ base clusters, and the same holds for~$B$, the short pair $(A,B)$ splits only into $O(1)$ well-separated pairs $\bigl(A\cap U_0(x),\,B\cap U_0(y)\bigr)$ in the partition WSPD, so the size bound of $O(n\eps^{-2}\log(1/\eps))$ still applies.

Finally, we bound the running time. At each node $A$ of the tree of the Euclidean WSPD
of bounding-box diagonal at most one, we store a list of centres of the base
clusters intersecting $A$. We build these lists bottom up. A leaf stores the
base centre of its vertex, and a parent merges its children's lists and
discards duplicates. By the packing argument above, each list has $O(1)$
entries, so all lists take time linear in the size of this tree to build. For each short pair $(A,B)$, combining its two lists gives $O(1)$
candidates for well-separated pairs. Each candidate for a well-separated pair is represented implicitly by $(A,x)$ and $(B,y)$.
If $x=y$ the candidate is output immediately. If $x \neq y$,
the pair of base clusters is marked exactly when the recursion in
\hyperref[sec:step2]{Step~2} reaches these two clusters and stops
without output. We store these unordered centre pairs in a list.
Using radix sorting on pairs of centre indices, we match the remaining
candidates against this list in time linear in $n$ plus the number
of candidates and listed pairs. We output a candidate exactly when
its centre pair appears in the list. There are $O(n\eps^{-2})$
candidates by Section~\ref{sec:step1} and
$O(n\eps^{-2}\log(1/\eps))$ listed pairs by Lemma~\ref{lem:calls}.
Therefore, the running time of $O(n\log n+n\eps^{-2}\log(1/\eps))$ still applies for partition WSPDs.
\end{proof}

\section{Distance oracle}
\label{sec:distance_oracle}

In this section, we extend the construction in Section~\ref{sec:construction} to an approximate distance oracle. The main task is construct a membership oracle, that is, a data structure that returns the well-separated pair containing two query vertices. Once we have a membership oracle, it is straightforward to retrieve an approximate distance for this well-separated pair, via a lookup table~\cite{gao-zhang,DBLP:conf/soda/GudmundssonW26}. We will use the word RAM model of computation~\cite{DBLP:journals/siamcomp/Har-PeledM06,LW21,DBLP:journals/jacm/Thorup04,DBLP:conf/stoc/ThorupZ01}, as is standard for $O(1)$-query time approximate distance oracles.

\subsection{Membership oracle}

Given a pair of distinct vertices $u$ and $v$, the membership oracle returns a well-separated pair $(A,B)$ from Section~\ref{sec:construction} satisfying $u \in A$ and $v \in B$, or $v \in A$ and $u \in B$. We will use the following three facts in our construction; all three facts hold under the word RAM model of computation.

\begin{fact}[Euclidean WSPD membership oracle, \cite{DBLP:conf/soda/GudmundssonW26}]
    \label{fact:euclidean_membership}
    Given $\varepsilon > 0$ and a set of $n$ points in~$\mathbb R^2$, one can construct a $(1/\varepsilon)$-WSPD and a membership oracle with $O(n \varepsilon^{-2})$ space, in $O(n \log n + n \varepsilon^{-2})$ preprocessing time, such that for any pair of points $(p,q)$ in the point set, the oracle returns the well-separated pair $(A,B)$ satisfying $p \in A$ and $q \in B$, or $q \in A$ and $p \in B$, in $O(1)$ time.
\end{fact}

\begin{fact}[Constant-factor approximate distance oracle for planar graphs, Theorem~3 in \cite{LW21}]
    \label{fact:planar_distance_oracle}
    Given an edge-weighted planar graph with $n$ vertices, one can construct an 8-approximate distance oracle in $O(n)$ space and $O(n \log^3 n)$ time, and can answer queries in $O(1)$ time.
\end{fact}

\begin{fact}[Level ancestor queries, ~\cite{DBLP:journals/tcs/BenderF04,DBLP:journals/jcss/BerkmanV94,DBLP:conf/wads/Dietz91}]
    \label{fact:level_ancestor_queries}
    Given a rooted tree~$T$ with $n$ nodes, one can preprocess the tree in $O(n)$ time and space, such that given a query node $u$ and a query depth $d$, one can return the ancestor of $u$ with depth $d$, in $O(1)$ time. 
\end{fact}

Next, we construct the membership oracle. 
For short pairs, we will use Fact~\ref{fact:euclidean_membership} to check the Euclidean well-separated pair. For long pairs, we will use Fact~\ref{fact:planar_distance_oracle} to estimate the distance between vertices, and then use Fact~\ref{fact:level_ancestor_queries} to search a constant number of net-tree levels.

\begin{lemma}
    \label{lem:membership_oracle}
    Given a connected Euclidean-weighted unit-disk graph on $n$ vertices in the plane, and $0 < \varepsilon \leq 1/8$, one can construct an $O(1)$-query time $(1/\varepsilon)$-WSPD membership oracle of $O(n \varepsilon^{-2} \log(1/\varepsilon)$ size in expected $O(n \log^3 n + n \varepsilon^{-2} \log(1/\varepsilon))$ processing time.
\end{lemma}

\begin{proof}
    First, we will describe the preprocessing procedure. Use Theorem~\ref{thm:main_introduction}(i) to construct a $(1/\varepsilon)$-WSPD, and store its short pairs and long pairs in (separate) lookup tables. 
    Then, construct three additional data structures.
    In Step~1, we use the WSPD of Fact~\ref{fact:euclidean_membership} with $1/s$ in place of $\varepsilon$, and keep its membership oracle. 
    Use Fact~\ref{fact:planar_distance_oracle} to construct a constant-factor approximate distance oracle for the planar 3-spanner from Step~2.
    Use Fact~\ref{fact:level_ancestor_queries} to preprocess the net-tree from Step~2 for level ancestor queries.

    Next, we will describe the query procedure. Given distinct vertices $u$ and $v$, query the Euclidean WSPD membership oracle (Fact~\ref{fact:euclidean_membership}) for the Euclidean well-separated pair~$(A,B)$ containing $(u,v)$. If $(A,B)$ is a short pair in the lookup table, return it. Otherwise, query the planar spanner approximate distance oracle (Fact~\ref{fact:planar_distance_oracle}) to obtain a 24-approximation of the distance between~$u$ and~$v$, let it be~$D \in [d_G(u,v), 24 \cdot d_G(u,v)]$. Loop over the net-tree levels~$i$ satisfying $r_i \in [\frac {\varepsilon D} {1200c^2}, \varepsilon D]$, where~$c$ is the constant from Fact~\ref{fact:LW}. Use the level ancestor query (Fact~\ref{fact:level_ancestor_queries}) to find the net-tree nodes $A_i$ and $B_i$ with level~$i$, and where $A_i$ is an ancestor of $u$ and $B_i$ is an ancestor of $v$. Check if $(A_i, B_i)$ is a long pair in the lookup table. If $i > 0$, also check $(A_{i-1},B_i)$ and $(A_i,B_{i-1})$. Return the first long pair found.

    We will prove the correctness of our algorithm. By definition, every returned pair contains~$(u,v)$ and belongs to the $(1/\varepsilon)$-WSPD in Theorem~\ref{thm:main_introduction}(i). Suppose the query does not return a short pair. By Lemma~\ref{lem:cover}, some long pair \((A,B)\) contains \(u\) and \(v\). Let \(d=d_G(u,v)\), and let \(i\) be the larger of the levels of \(A\) and \(B\). Since the recursion stopped at the call \((A,B)\), by  Lemma~\ref{lem:test} and Lemma~\ref{lem:hierarchy}(ii), we have \(d>\frac{8cr_i} {\varepsilon}-4cr_i\ge \frac{4cr_i} {\varepsilon}\). Since the recursion did not stop at the previous call, by Lemmas~\ref{lem:call-distance} and Lemma~\ref{lem:hierarchy}(ii), we have \(d\le \frac{49c^2r_i}{\varepsilon}+4cr_i\le \frac{50c^2r_i}{\varepsilon}\). Rearranging these inequalities, we get
    $r_i\in\left[\frac{\varepsilon d}{50c^2},\frac{\varepsilon d}{4c}\right]$, which implies
    $r_i \in \left[\frac{\varepsilon D}{1200c^2},\varepsilon D\right]$. Thus the query examines level \(i\). By Lemma~\ref{lem:levels}, the other cluster has level \(i\) or \(i-1\), so \((A,B)\) is one of the pairs that we checked, as required.

    Finally, we analyse the size, preprocessing time, and query time. By Theorem~\ref{thm:main_introduction}(i), the size of the short and long pair lookup tables is $O(n\varepsilon^{-2} \log(1/\varepsilon)$, whereas the additional data structures require $O(n \varepsilon^{-2})$ space by
    Fact~\ref{fact:euclidean_membership}, Fact~\ref{fact:planar_distance_oracle} and Fact~\ref{fact:level_ancestor_queries}. The processing time is also dominated by constructing the WSPD in $O(n\log n+n\eps^{-2}\log(1/\eps))$ time (Theorem~\ref{thm:main_introduction}(i)), and constructing the planar distance oracle in $O(n\log^3 n)$ time (Fact~\ref{fact:planar_distance_oracle}).
    The total expected preprocessing time is therefore $O(n\log^3 n+n\eps^{-2}\log(1/\eps))$. The query time for short pairs is $O(1)$ by Fact~\ref{fact:euclidean_membership}. For long pairs, computing~$D$ takes $O(1)$ time by Fact~\ref{fact:planar_distance_oracle}, and checking the $O(1)$ levels in the net-tree takes $O(1)$ time by Fact~\ref{fact:level_ancestor_queries}. Thus, the query time is $O(1)$.
\end{proof}

\subsection{Distance oracle}

We can now build a distance oracle in essentially the same way as~\cite{gao-zhang} and~\cite{DBLP:conf/soda/GudmundssonW26}.
This leads to \autoref{thm:oracle}, which we restate here for convenience.

\DistanceOracle*

\begin{proof}
    At preprocessing time, construct an $(8/\eps)$-WSPD membership oracle using Lemma~\ref{lem:membership_oracle}. For each WSPD pair $(A,B)$, choose representatives $a\in A$ and $b\in B$. Using the distance oracle of Chan and Skrepetos~\cite[Corollary~2]{chan-skrepetos}, compute a $(1+\eps/8)$-approximation $d'_G(a,b)$ and store $w(A,B)=(1+\eps/4)d'_G(a,b)$. Then discard their oracle. At query time, for a query pair $(u,v)$, use the membership oracle to find a WSPD pair $(A,B)$ containing $(u,v)$ and return $w(A,B)$.

    Next, we prove correctness. It is straightforward to verify that all pairs in a $(1/\delta)$-WSPD have a pairwise distance within $(1+2 \delta)$ of one another. Let $(A,B)$ be the well-separated pair containing $(u,v)$ and let the representatives of $(A,B)$ at preprocessing time be $(a,b)$. Then,
    \begin{align*}
        d_G(u,v) &\leq (1+\varepsilon/4) \cdot d_G(a,b) \leq (1+\varepsilon/4) \cdot d'_G(a,b) = w(A,B),
        \\
        w(A,B) &= (1+\varepsilon/4) \cdot d'_G(a,b) \leq (1+\varepsilon/4) \cdot(1+\varepsilon/8) \cdot d_G(a,b)
        \\
        &\leq (1+\varepsilon/4)^2 \cdot(1+\varepsilon/8) \cdot d_G(u,v) \leq (1+\varepsilon) \cdot d_G(u,v).
    \end{align*}
    Therefore, $w(A,B)$ is a $(1+\varepsilon)$-approximation of $d_G(u,v)$, as required.

    The size is $O(n\varepsilon^{-2} \log(1/\varepsilon)$, from the size of our WSPD (Theorem~\ref{thm:main_introduction}(i))
    and our membership oracle (Lemma~\ref{lem:membership_oracle}). 
    The expected preprocessing time is $O(n\eps^{-5}\log^3 n+n\eps^{-6}\log(1/\eps))$: we construct the Chan and Skrepetos~\cite{chan-skrepetos} distance oracle and query it once per WSPD pair. The query time is $O(1)$: it only requires a membership query and a lookup of $w(\cdot, \cdot)$.
\end{proof}

\section{AI Statement}

The results were developed through interactions between the authors and LLMs (ChatGPT 6 and Claude 5.5). The paper was written by the authors with assistance from LLMs.

\bibliography{bib} 

\end{document}